\documentclass[aps,prd,12pt,nofootinbib,floatfix]{revtex4}
\usepackage{epsfig}
\usepackage{graphicx}
\usepackage[font=small,skip=0pt,justification=raggedright]{caption}
\usepackage{amsmath}
\usepackage{amssymb}
\usepackage{mathrsfs}
\usepackage{verbatim}
\usepackage{dutchcal}

\usepackage[normalem]{ulem}
\usepackage{xcolor}

\newcounter{fig}   \newcommand{\lbfig}[1]{\refstepcounter{fig}
\label{#1} }

\newcommand{\bea}{\begin{eqnarray}}
\newcommand{\eea}{\end{eqnarray}}
\newcommand{\be}{\begin{equation}}
\newcommand{\ee}{\end{equation}}

\newcommand{\re}[1]{(\ref{#1})}

\newcommand{\cgg}{\mathcal{g}}

\newcommand{\eqn}{\begin{eqnarray}}
\newcommand{\eqnx}{\end{eqnarray}}

\begin{document}

\title{\boldmath Two types of boson stars\\ in  the  
$3+1$-dimensional $O(3)$ sigma-model \unboldmath}
\author{Jutta Kunz}
\email{jutta.kunz@uni-oldenburg.de}
\affiliation{Institute of Physics,
Carl von Ossietzky University Oldenburg, Germany
Oldenburg D-26111, Germany}

\author{Aliaksei Mikhaliuk}
\email{alekseybsu.mihalyuk@gmail.com}
\affiliation{Belarusian State University, Minsk 220004, Belarus}

\author{Yakov Shnir}
\email{shnir@theor.jinr.ru}
\affiliation{BLTP, JINR, Dubna 141980, Moscow Region, Russia}

\date{\today}

\begin{abstract}
We investigate boson stars in an $O(3)$ scalar field theory with a symmetry-breaking potential.
By constructing numerically spherically symmetric solutions, we demonstrate that the model gives rise to a rich set of field configurations.
The negative coupling constant of the scalar self-interactions allows for two types of boson stars:
Type I solutions represent the usual boson stars, that emerge from the vacuum as the boson frequency is decreased below the boson mass, whereas type II boson stars emerge from a set of static soliton solutions.
Depending on the strength of the gravitational coupling constant, both types or only one type is present.
At a critical set of coupling constant, both types undergo a bifurcation. 
There the spirals of both types disconnect from their branches and reconnect with each other, while the remaining branches of both types also reconnect with each other.
Analogous features are seen for U(1) gauged boson stars.

\end{abstract}

\maketitle

\newpage

\section{Introduction}
As is well-known, minimal coupling of Einstein gravity to a massive scalar theory gives rise to macroscopic particle-like localized  configurations dubbed as boson stars \cite{Kaup:1968zz,Feinblum:1968nwc,Ruffini:1969qy,Colpi:1986ye}, see  \cite{Jetzer:1991jr,Schunck:2003kk,Liebling:2012fv,Shnir:2022lba} for detailed reviews. 
Boson stars represent stationary solutions of the coupled Einstein-scalar field equations, when the complex scalar field has a harmonic time dependence. 
The boson stars in the Einstein-Klein-Gordon theory without a self-interaction potential do not possess a flat space limit. 
When a self-interaction is included, a resulting family of more massive boson stars arises \cite{Colpi:1986ye,Friedberg:1986tq}.
In the model with a sextic potential \cite{Friedberg:1986tq,Kleihaus:2005me} these boson stars tend to the corresponding non-topological solitons, the Q-balls \cite{Rosen:1968mfz,Friedberg:1976me,Coleman:1985ki}, in the flat space limit. 

Both Q-balls and boson stars are regular stationary solutions.
The phase invariance of the scalar field is associated with a conserved global Noether charge, which is proportional to the angular frequency $\omega$ of the scalar field.
Typically, a family of boson stars exists for a certain range of values of the angular frequency between a maximal value $\omega_{max}$, which corresponds to the mass of the scalar excitations, and some lower nonzero critical value $\omega_{min}$, that depends on the properties of the potential.

The global $U(1)$ symmetry of a scalar model supporting Q-balls can be promoted to a local gauge symmetry.
This gives rise to gauged solitons possessing electric charge \cite{Rosen:1968mfz,Lee:1988ag,Anagnostopoulos:2001dh,Gulamov:2013cra,Gulamov:2015fya,Kunz:2021mbm}.
The presence of the long-range gauge field affects the properties of the solitons.
A strong electromagnetic repulsion then reduces the domain of existence of charged Q-balls and charged boson stars.  
The gravitational attraction, however, produces an opposite effect.
Instead of the paradigmatic spiraling scenario of uncharged boson stars, one finds a more complicated pattern of dynamical evolution for $U(1)$ gauged boson stars \cite{Kunz:2021mbm}.

The non-linear $O(3)$ sigma model with a non-negative potential also supports boson stars \cite{Verbin:2007fa,Herdeiro:2018djx,Cano:2023bpe,Adam:2025ktm}. 
These oscillate or rotate in internal space, and do not possess flat space counterparts. 
This is similar to the mini-boson stars in the Einstein-Klein-Gordon theory.
They arise from perturbative excitations of the vacuum at $\omega_{max}$, where both the mass and the Noether charge trivialize. 
The $O(3)$ boson stars exhibit the typical spiraling pattern of the frequency dependence of both charge and mass \cite{Herdeiro:2018djx}.

On the other hand, static regular soliton solutions exist in the $O(3)$ sigma model with a modified symmetry breaking potential in Minkowski space \cite{Ferreira:2025xey}. 
Since this potential is partly negative, it allows to circumvent Derrick's theorem (see e.g.~\cite{Nucamendi:1995ex,Bechmann:1995sa,Khlopov:1985fch,Kleihaus:2013tba,Chew:2024bec}).
Notably, such soliton configurations also admit oscillations or rotations in internal space. 
Hence one may expect the existence of a different family of $O(3)$ boson stars, which are linked to the self-gravitating non-topological static lumps in Minkowski spacetime in the limit $\omega=0$.  

The main purpose of this paper is to report the existence of such a new type (type II) of boson stars, which emerge in the $O(3)$ non-linear sigma-model minimally coupled to Einstein gravity. 
We here construct spherically symmetric solutions and investigate their physical properties for a choice of the scalar potential which, apart from the usual ``pion mass'' term, also contains a non-positive term which yields an additional short-range interaction energy \cite{Ferreira:2025xey} (see also 
\cite{Leese:1989gi,Salmi:2014hsa,Samoilenka:2015bsf,Gillard:2015eia}).
Since these type II boson stars owe their existence to the presence of the symmetry breaking potential in the presence of a negative energy contribution, they were not feasible in the previous study \cite{Herdeiro:2018djx}.
We first address these solutions in the ungauged model and then consider the influence of electrostatic repulsion. 

The model also supports spherically symmetric solutions of a different type that are analogous to the usual boson stars (type I).
These configurations with a time-dependent scalar phase emerge smoothly from linearized perturbations around Minkowski spacetime as $\omega$ approaches the mass threshold. 
These type I boson stars persist, when the the symmetry breaking potential is switched off \cite{Herdeiro:2018djx}.
We observe that, as long as the effective gravitational coupling is relatively strong, both types of boson stars show a spiraling behavior for their masses and charges towards two different limiting solutions at the centers of the spirals at the frequencies $\omega^{(I)}_{cr} > \omega^{(II)}_{cr}$, respectively. 

However, as the gravitational coupling becomes weaker, the pattern exhibited in the mass-frequency diagram becomes very different:
boson stars exist for the whole range of values of the angular frequency $\omega \in [0:\omega_{max}]$.
First, the spirals only overlap, and then the type I and type II solutions bifurcate at a critical value of the coupling.
This intriguing phenomenon has not been observed before.
The presence of an electric charge further affects the force balance between the gravitational attraction and the short- and long-range scalar interactions by providing in addition an electrostatic repulsion.
Consequently, the bifurcations between the two different types of boson stars occur at critical values of the gauge coupling that depend on the value of the gravitational coupling.

This paper is organized as follows. 
In Sec.~II, we introduce the model and the field equations with the total stress-energy tensor of the system of interacting fields. 
In Sec.~III we describe the spherically symmetric parametrization of the metric and the matter fields.
Here we also discuss the physical quantities of interest and the appropriate set of boundary conditions. 
The numerical results are presented in Sec.~IV.
First, we discuss the domain of existence and the properties of the ungauged spherically symmetric boson stars of both types and discuss their dependence on the parameters of the model.
Next, the properties of $U(1)$ gauged boson stars are analyzed.
Finally, in Sec.~V we present some conclusions and further remarks.

\section{The model and field equations}

We consider the (3+1)-dimensional action 
\be
S=\int d^4 x \sqrt{-{\mathcal g}} \left(\frac{R}{16\pi G} - L_{m}\right)\, .
\label{lag}
\ee
Here the gravity part is the usual Einstein-Hilbert action, $R$ is the Ricci scalar associated with the spacetime metric $g_{\mu\nu}$ with the determinant ${\mathcal g}$, and $G$ is Newton's constant. The signature of the Minkowski spacetime metric is chosen to be (-, +, +, +).
The Lagrangian of the matter fields $L_m$ is given by the $U(1)$ gauged non-linear sigma-model
\be
L_{m} = \frac14 F_{\mu\nu}F^{\mu\nu} + \frac12 (D_\mu\phi^a)^2 + U(\phi) \, .
\label{Gauged_O3}
\ee
Here the real triplet of the scalar fields $\phi^a$, $a=1,2,3$, is restricted to the surface of the unit sphere, $(\phi^a)^2=1$, and $U(\phi)$ is a symmetry breaking potential. 
The $U(1)$ field strength tensor is $F_{\mu\nu}=\partial_\mu A_\nu-\partial_\nu A_\mu$, and the covariant derivative of the field $\phi^a$ 
is
\be
D_\mu \phi^\alpha = \partial_\mu \phi^\alpha
+e\,A_\mu\,\varepsilon_{\alpha\beta}\,\phi^\beta \,, \qquad D_\mu
\phi^3 = \partial_\mu \phi^3, \qquad \alpha,\,\beta=1,\,2 \,
\label{covariant}
\ee
with the gauge coupling  $e$.  

Following the previous paper 
\cite{Ferreira:2025xey}, 
we consider a symmetry breaking potential 
\be
U(\phi)=\mu^2 (1 -\phi^3) - \beta (1-\phi^3)^4 \,  ,
\label{pot-mod}
\ee
where $\mu$ and $\beta $ are real positive parameters. 
We note, that the potential breaks explicitly the
original O(3) symmetry of the kinetic term, but the U(1) symmetry stays unbroken.
Hence, the Lagrangian \re{Gauged_O3} is invariant with respect to the local Abelian gauge transformations
\be
(\phi^1 + i \phi^2) \to
e^{ie \zeta}(\phi^1 + i \phi^2), \quad A_\mu \to A_\mu - \partial_\mu \zeta  \, ,
\label{rotate}
\ee
where $\zeta$ is a real function of the coordinates.
We note that one can set $\mu=1$ via rescaling of the parameter $\beta \to \beta/\mu^2$. 
Further, rescaling of the coordinates $x_\nu\to \mu x_\nu$ yields the effective gravitational coupling $\alpha^2=8\pi G \mu^2$.

Importantly, the potential \re{pot-mod}, apart from the weakly attractive ``pion mass'' term, also contains a term which yields an additional short-range interaction \cite{Leese:1989gi,Salmi:2014hsa,Gillard:2015eia}. 
Potentials of that type support localized non-topological soliton solutions in 3+1-dimensional flat spacetime \cite{Verbin:2007fa,Ferreira:2025xey}. When not being positive everywhere, these potentials allow circumvention of Derrick's theorem and give rise to solitons often dubbed scalarons\footnote{Such solitons should not be confused with the scalarons introduced by Starobinsky in a cosmological context \cite{Starobinsky:1980te}.} (see e.g.~\cite{Nucamendi:1995ex,Kleihaus:2013tba,Chew:2024bec}).
Moreover, the potential \re{pot-mod}  allows for the presence of a harmonic time dependence of the scalar field, which is a key property of various flat space Q-balls \cite{Rosen:1968mfz,Friedberg:1976me,Coleman:1985ki}. 

Variation of the action \re{lag} with respect to the metric leads to the Einstein equations
\be
R_{\mu\nu} -\frac12 R g_{\mu\nu} = 2\alpha^2 \left( T_{\mu\nu}^{Em} + T_{\mu\nu}^{\phi}\right) \, ,
\label{einsteq}
\ee
where the electromagnetic and scalar components of the energy-momentum tensor are
\be
\begin{split}
T_{\mu\nu}^{Em} &=F_\mu^\rho F_{\nu\rho} - \frac14 g_{\mu\nu} F_{\rho\sigma}  F^{\rho\sigma}\,,\\
T_{\mu\nu}^{\phi} &= D_\mu\phi^a D_\nu\phi^a
 - g_{\mu\nu}\left[\frac{g^{\rho\sigma}}{2}  D_\rho\phi^a
D_\sigma\phi^a +  U(\phi)\right] \, .
\label{Teng}
\end{split}
\ee
The corresponding vector and scalar field equations are
\be
    \partial_\mu\left(\sqrt{-{\mathcal g}}  F^{\mu\nu}\right)=e \sqrt{-{\mathcal g}} j^\nu, \quad D^\mu D_\mu \phi^a + \phi^a( D^\mu \phi^b \, D_\mu \phi^b)+ \frac{\delta U}{\delta \phi}\phi^a =0\, ,
\label{eqfield}
\ee
where 
\be
j^\mu=\phi^1 D^\mu\phi^2 - \phi^2 D^\mu\phi^1
\label{ncurrent}
\ee
is the conserved Noether $U(1)$ current associated with the local $U(1)$ symmetry \re{rotate}. 
The corresponding charge is given by $Q=\int d^3 x \sqrt{-{\mathcal g}} j^0$ and represents the particle number. 

\section{Spherically symmetric ansatz and boundary conditions}

We are interested in stationary spherically symmetric regular solitonic
solutions of the system \re{einsteq}  -  \re{ncurrent}. 
To construct such solutions numerically we employ the line element \cite{Kunz:2023qfg,Herdeiro:2024yqa,Kunz:2024uux}
\be
ds^2=g_{\mu\nu}dx^\mu dx^\nu= -F_0(r)dt^2 +F_1(r)(dr^2 + r^2 d\Omega^2 ) \, ,
\label{metric}
\ee
where $d\Omega^2= d\theta^2 + \sin^2\theta d\varphi^2$, and the metric functions $F_0$ and $F_1$ depend on the isotropic radial coordinate $r$ only.
For the scalar $O(3)$ field, we adopt the ``trigonometric'' stationary ansatz 
\be
    \phi^a=[\sin{f(r)}\cos{(\omega t)},\text{ }\sin{f(r)}\sin{(\omega t)},\text{ }\cos{f(r)}] \, .
    \label{anssfield}
\ee
Further, in the static gauge the gauge potential can be written as
\be
A_\mu dx^\mu= A_0(r) dt \, .
\label{At}
\ee

The $U(1)$ gauged  self-gravitating regular configurations  are characterized by the Arnowitt-Deser-Misner (ADM) mass $M$ and the electric charge $Q_e$. 
These quantities can be extracted from the asymptotic behavior of the metric and the gauge field functions,
\be
g_{00}\rightarrow -1+\frac{\alpha^2 M}{\pi r}+O(\frac{1}{r^2}),\quad
A_0  \rightarrow \frac{Q_e}{r} +O(\frac{1}{r^2}) \, .
\ee
These quantities can also be computed as the integrals of the corresponding densities of the total  stress-energy tensor $T_{\mu\nu}=T^\phi_{\mu\nu} + T^{Em}_{\mu\nu}$, yielding the mass $M$,
\be
M=\int\sqrt{-\cgg} dr d\theta d\varphi\left( T^\mu_\mu - 2 T^0_0\right) \, ,
\label{mass-int}
\ee
and the Noether current \re{ncurrent} multiplied by the charge, yielding the electric charge $Q_e$,
\be
Q_e =e \int d^3 x \, \sqrt{-\cgg} \, j^0 
=8  \pi\, e \int_{0}^\infty dr\, r^2 \frac{F_1^{3/2}}{\sqrt{F_0 }}
 (\omega - e A_0) \sin^2 f
\, .
\label{chargeN}
\ee
The regular solutions satisfy the truncated first law of thermodynamics without the entropy term, $dM=\omega dQ$ (see e.g.~\cite{Herdeiro:2021mol,Stotyn:2013spa}). 
We used these relations as a test of the numerical accuracy.

The full system of the field equations \re{einsteq} -  \re{ncurrent}. 
can be solved numerically using the parametrization \re{metric} - \re{At}.
The corresponding boundary conditions are found by considering the asymptotic expansions of the equations on the boundary of the domain of integration together with the assumption of regularity and asymptotic flatness. 
Explicitly, we impose 
\begin{itemize}
    \item at $r=0:\qquad \partial_r f =  \partial_r A_0 = \partial_r F_0=\partial_r F_1 =0$ \ ,
    \item at $r=\infty:\qquad f=0,\quad \quad A_0=0,\quad F_0=F_1=1$ \ .
\end{itemize}

The solutions are found by using a sixth-order finite difference scheme.
The system of four second-order coupled ordinary differential equations is discretized on a grid with a typical size of 230 points in the radial direction. 
The emerging system of nonlinear algebraic equations has been solved using the professional package FIDISOL/CADSOL \cite{schoen} which uses a Newton-Raphson method, and with 
the Intel MKL PARDISO sparse direct solver \cite{MKL} and the CESDSOL library\footnote{Complex Equations-Simple Domain partial differential equations SOLver, a C++ package developed by I.~Perapechka, see, e.g., Refs. \cite{Herdeiro:2021mol,Kunz:2019sgn,Herdeiro:2021jgc}}. In all  cases, the typical errors are of order of $10^{-6}$.

\section{Solutions}

\subsection{Ungauged $O(3)$ boson stars}

\begin{figure}[t!]
\begin{center}
\includegraphics[height=.45\textheight,  angle =-0]{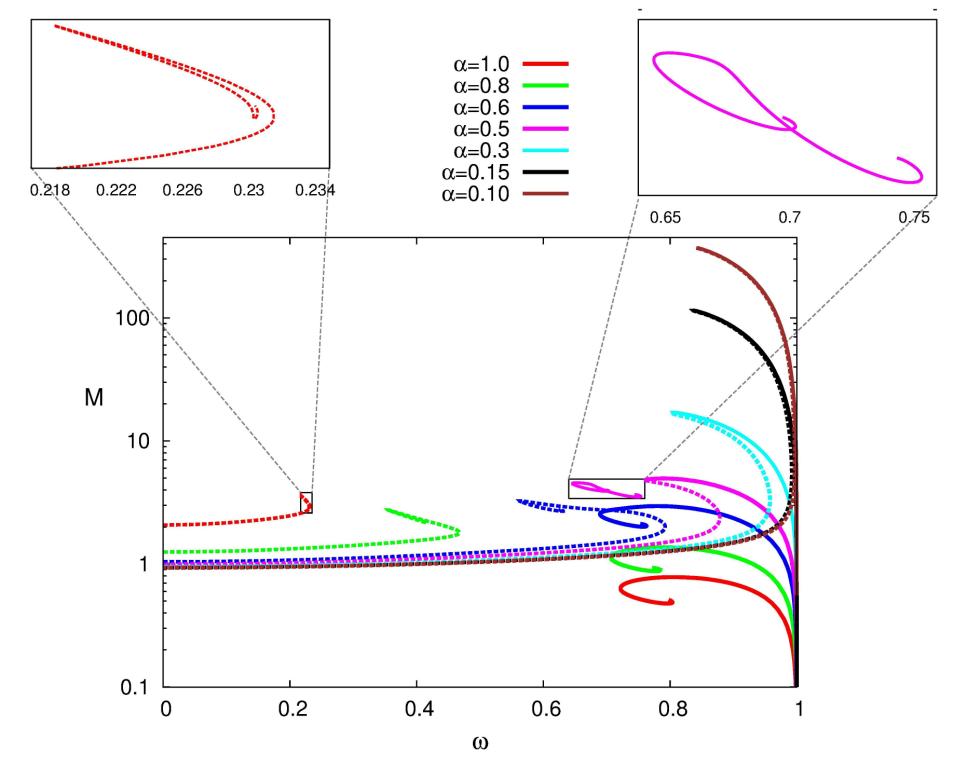}
\\[4pt]
\includegraphics[height=.36\textheight,  angle =-90]{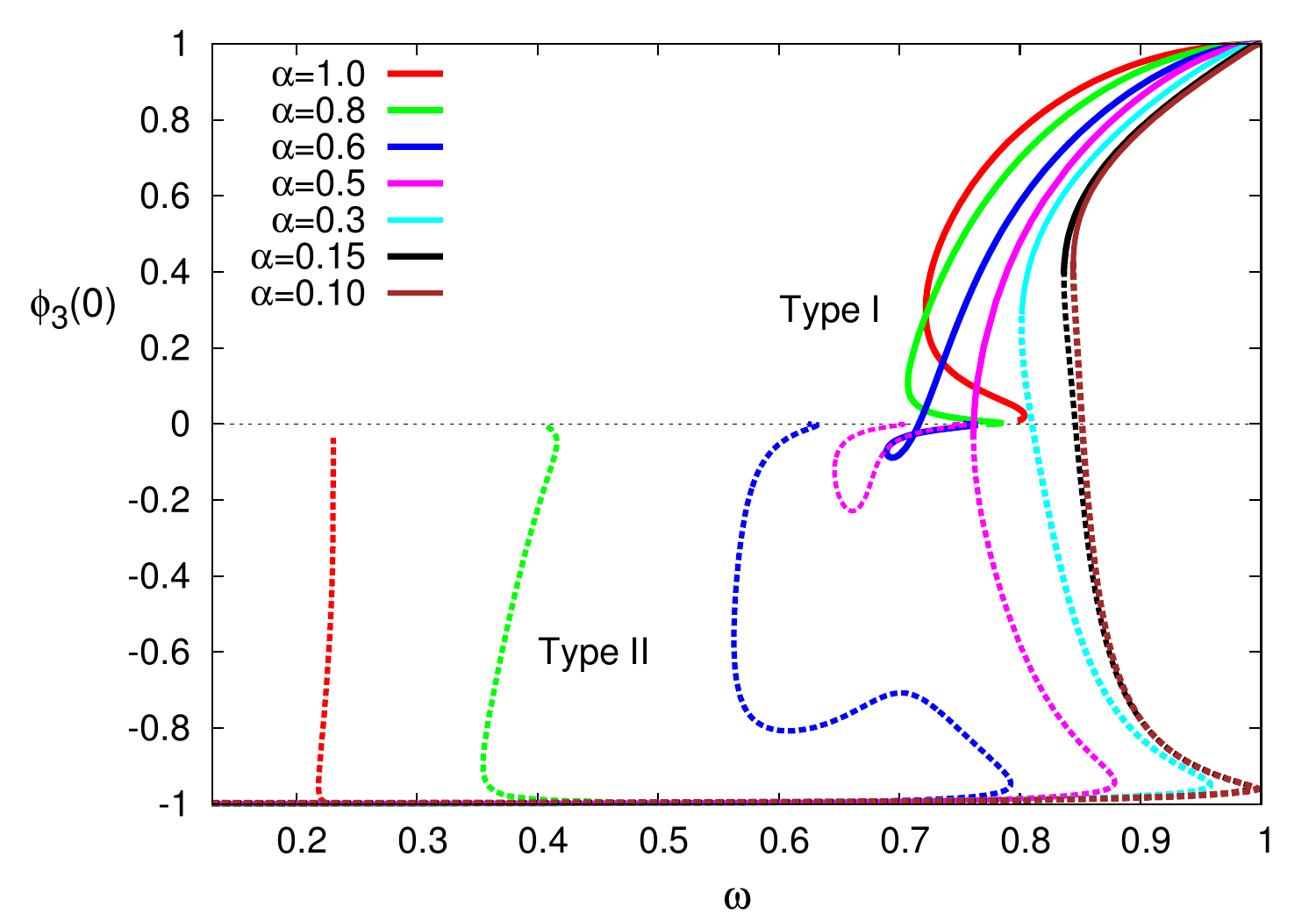}
\includegraphics[height=.36\textheight,  angle =-90]{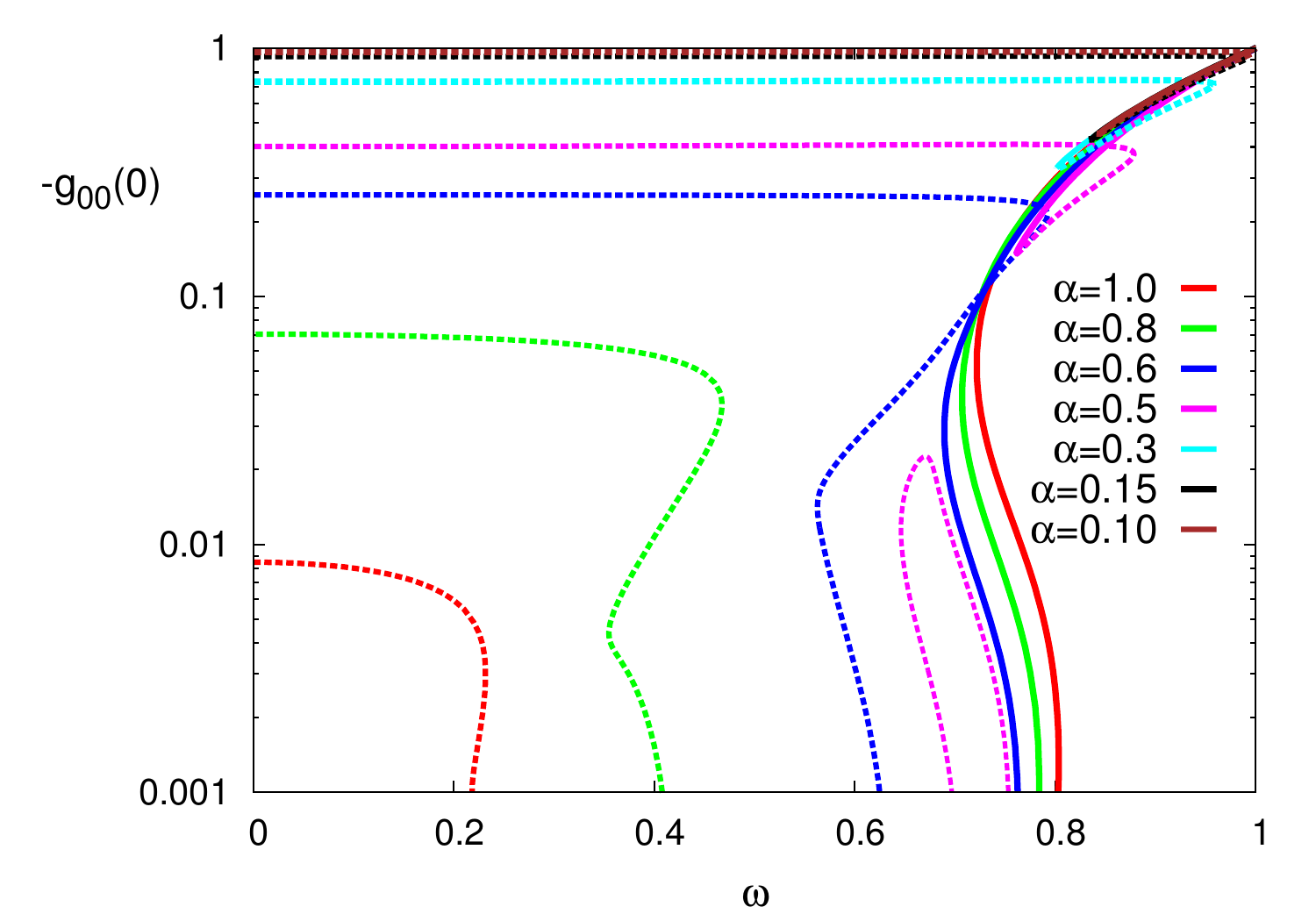}
\end{center}
\caption{\small Spherically symmetric ungauged ($e=0$) $O(3)$ boson stars: 
The ADM mass $M$ in units of $8\pi$ (upper plot), the central values of the scalar component $\phi_3(0)$,(bottom left) and the metric component $-g_{00}(0)$ (bottom right) vs the frequency $\omega$ are plotted for some set of values of the gravitational coupling $\alpha$ and $\beta=0.5$. 
The solid and dashed lines correspond to solutions of types I and II, respectively.}    
\lbfig{fig1}
\end{figure}

\begin{figure}[t!]
\begin{center}
\includegraphics[height=.38\textheight,  angle =-90]{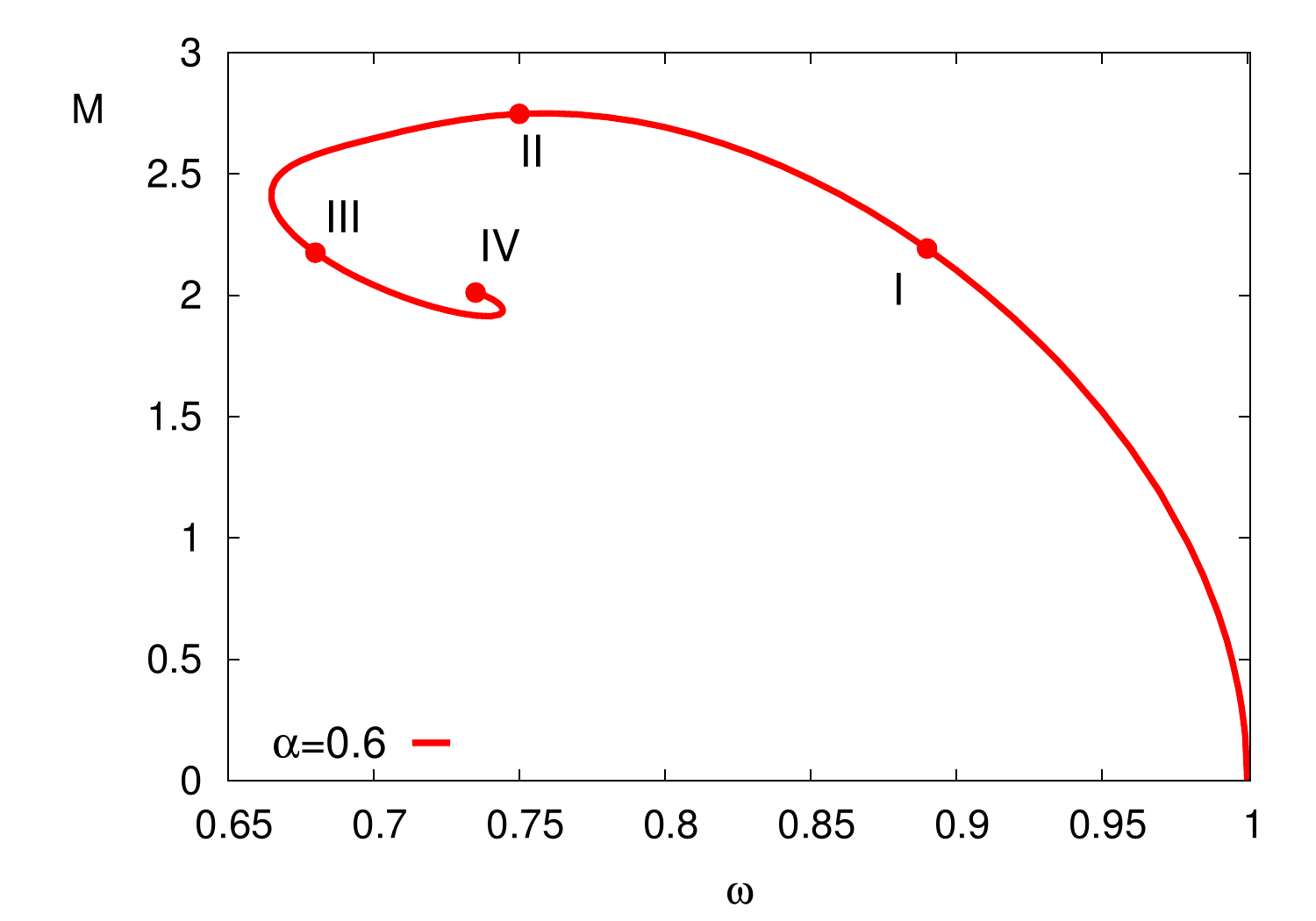}\\
\includegraphics[height=.32\textheight,  angle =-90]{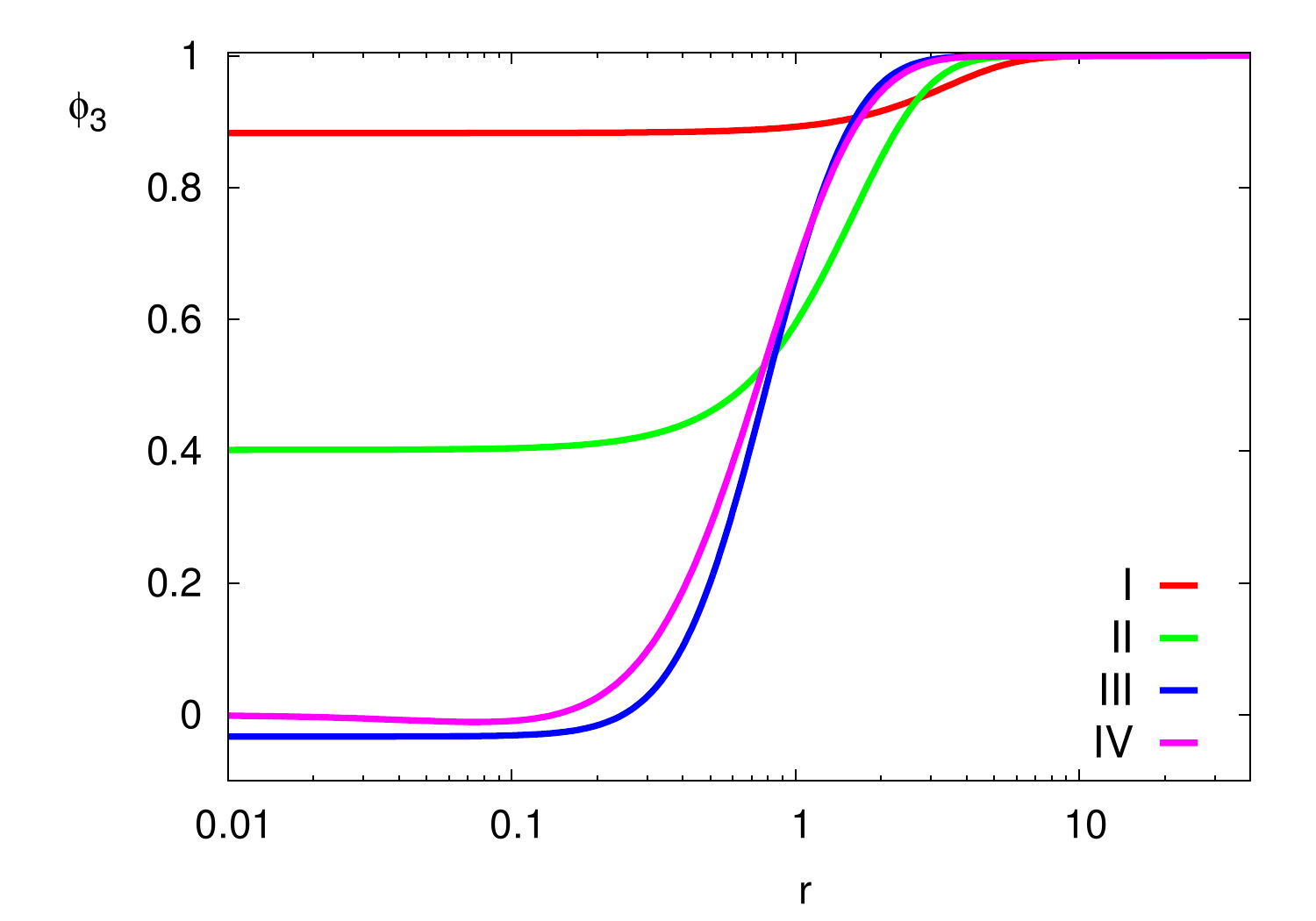}
\includegraphics[height=.32\textheight,  angle =-90]{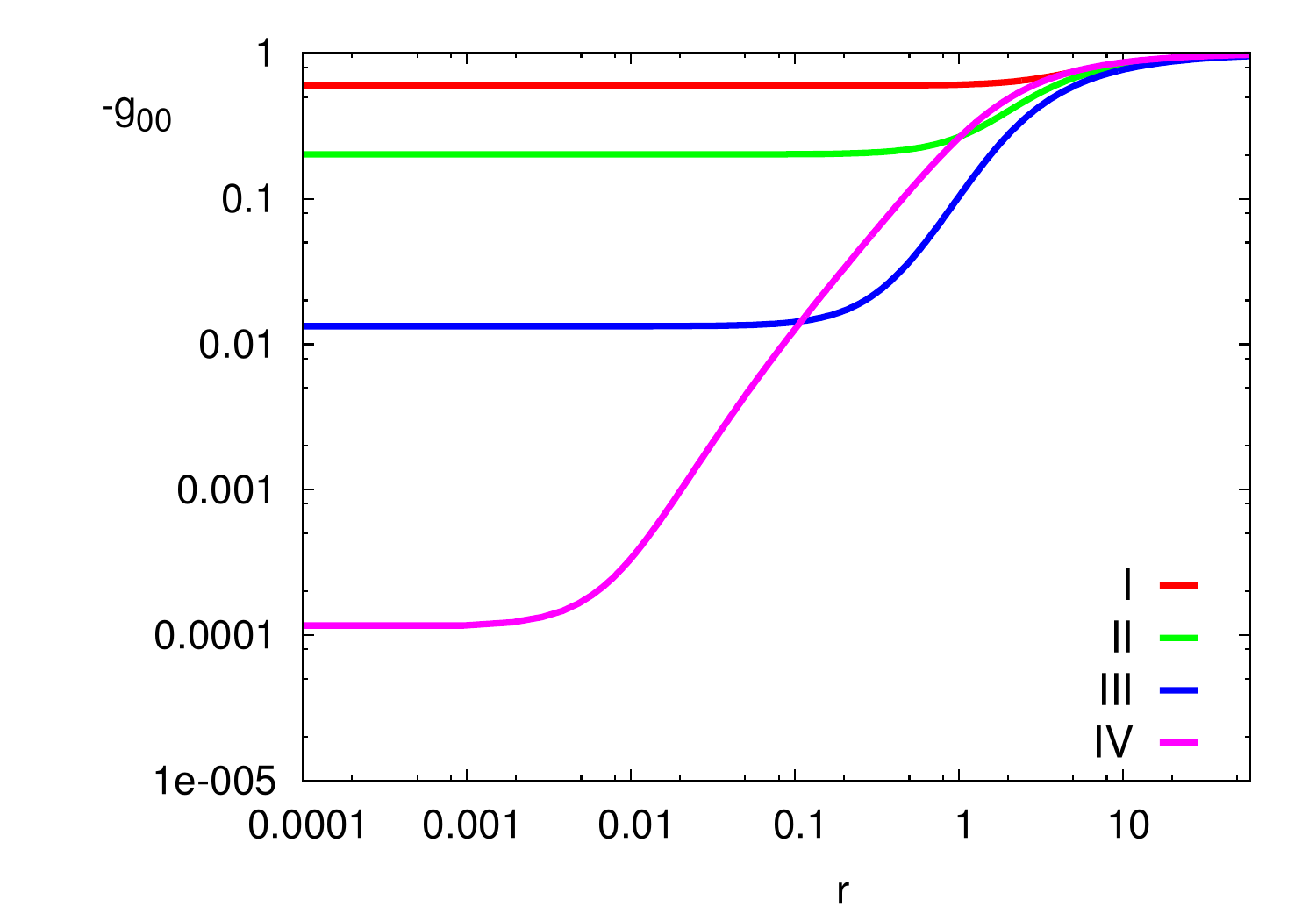}
\end{center}
\caption{\small Spherically symmetric ungauged ($e=0$) $O(3)$ boson stars: 
The ADM mass $M$ of the type I boson stars in units of $8\pi$ vs the frequency $\omega$ for $\alpha=0.6$ and $\beta=0.5$ (upper plot) and illustrative radial profiles of the scalar component $\phi_3$ (bottom left), and the metric component $-g_{00}$ (bottom right) of the solutions, indicated by dots on the upper plot.}    
\lbfig{fig2}
\end{figure}

\begin{figure}[t!]
\begin{center}
\includegraphics[height=.38\textheight,  angle =-90]{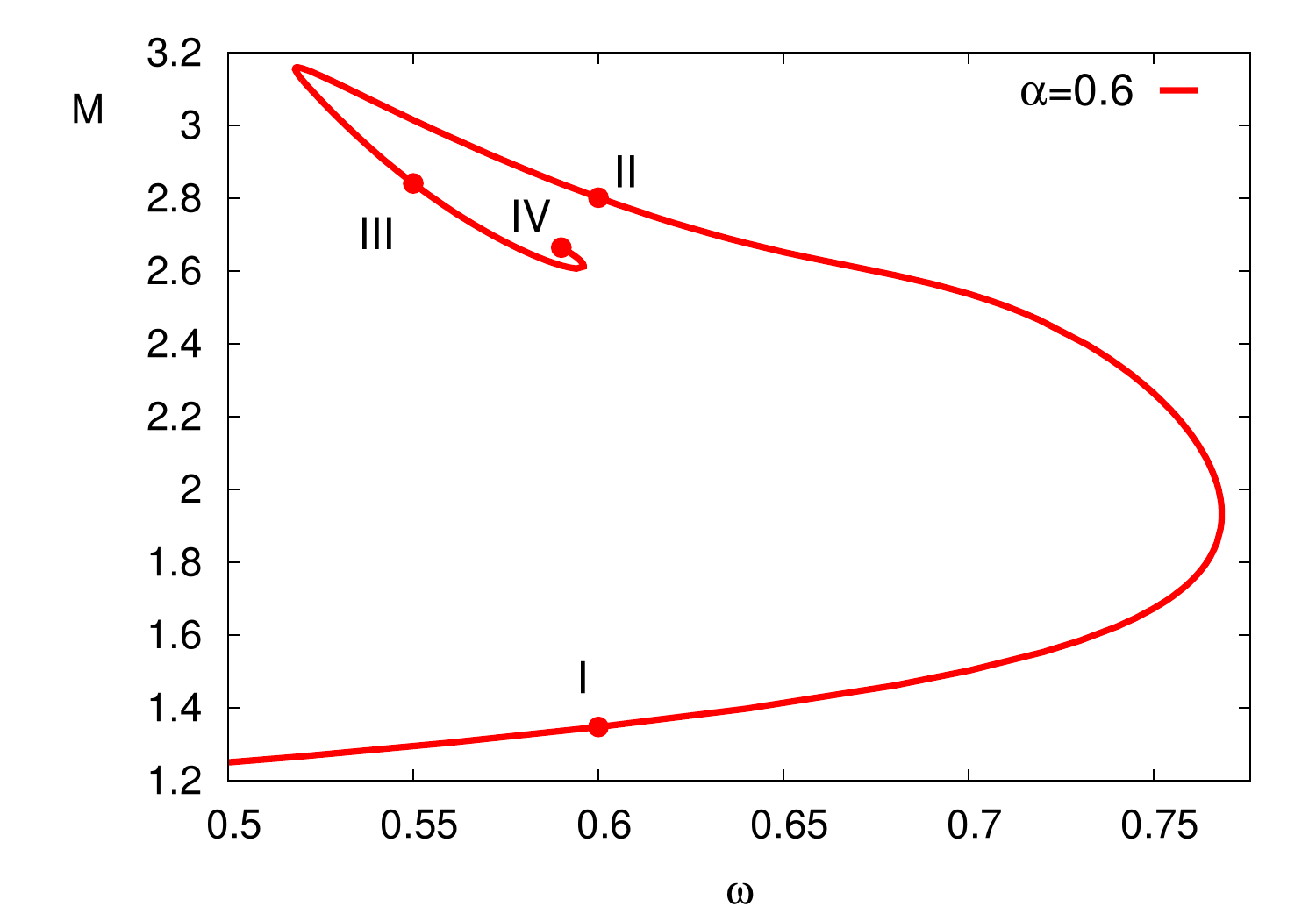}\\
\includegraphics[height=.32\textheight,  angle =-90]{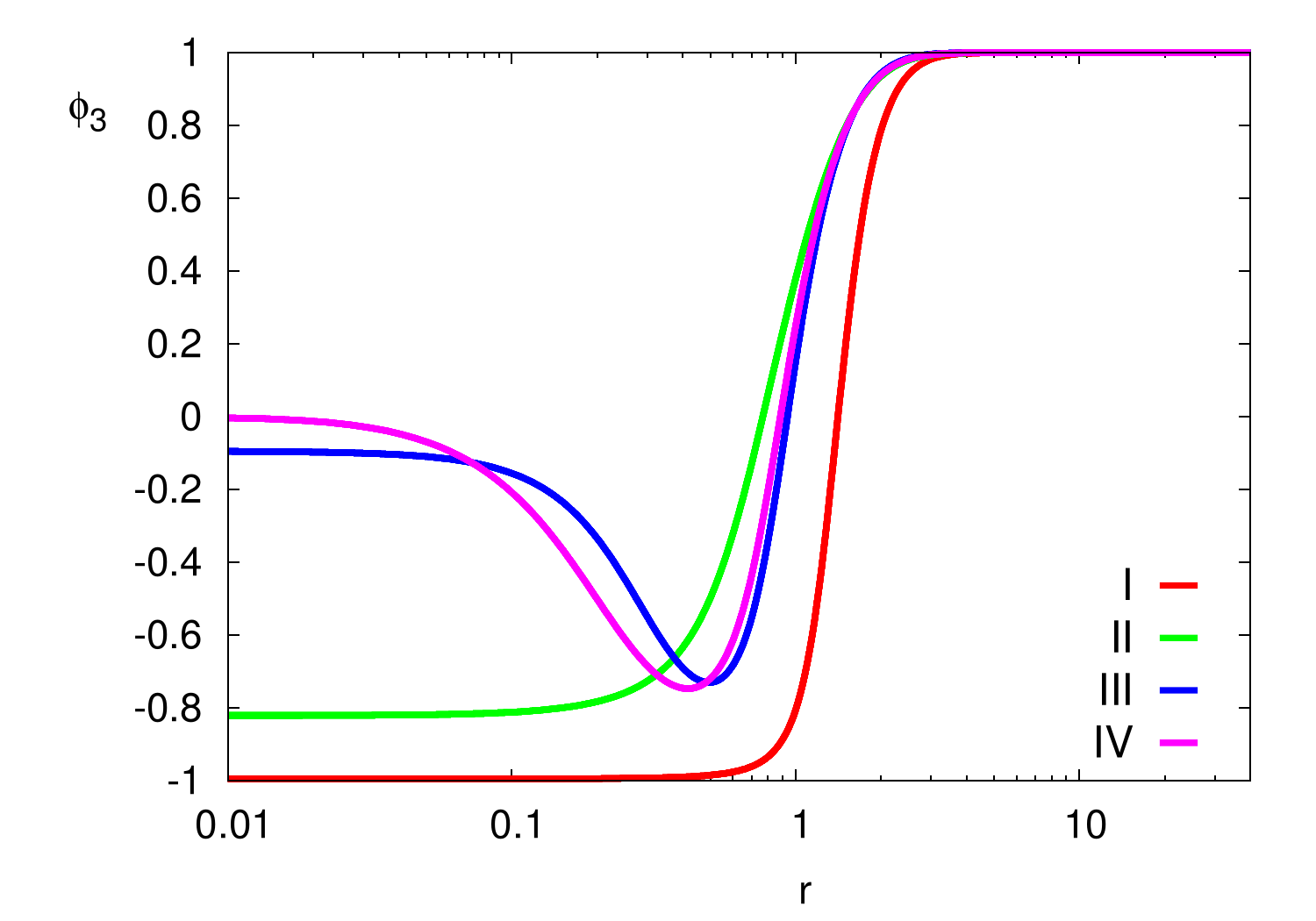}
\includegraphics[height=.32\textheight,  angle =-90]{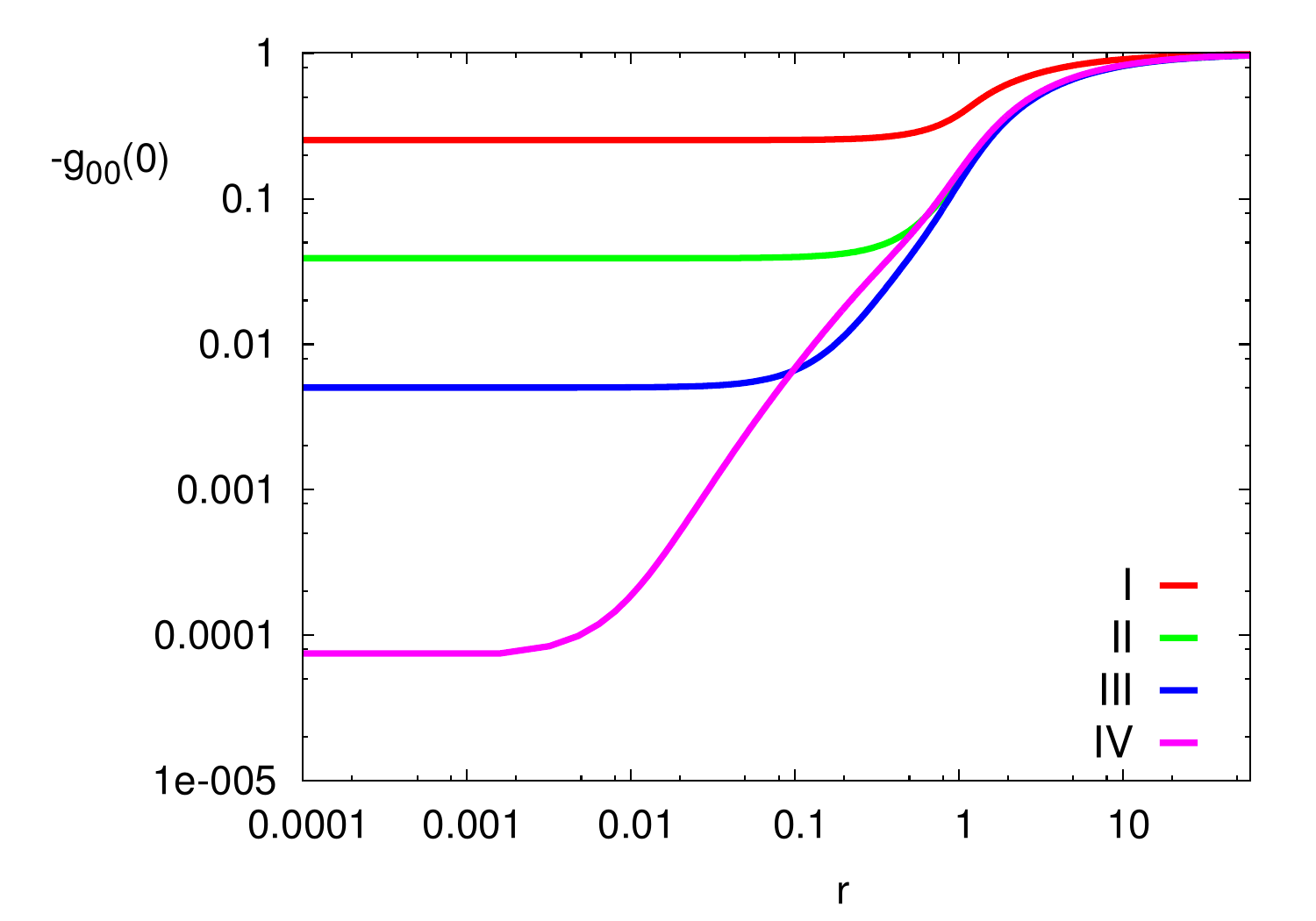}
\end{center}
\caption{\small Spherically symmetric ungauged ($e=0$) $O(3)$ boson stars: 
The ADM mass of the type II boson stars in units of $8\pi$ vs the frequency $\omega$ for $\alpha=0.6$ and $\beta=0.5$ (upper plot) and illustrative radial profiles of the scalar component $\phi_3$ (bottom left) and the metric component $-g_{00}$ (bottom right) of the solutions, indicated by dots on the upper plot.}    
\lbfig{fig3}
\end{figure}

\begin{figure}[t!]
\begin{center}
\includegraphics[height=.38\textheight,  angle =-90]{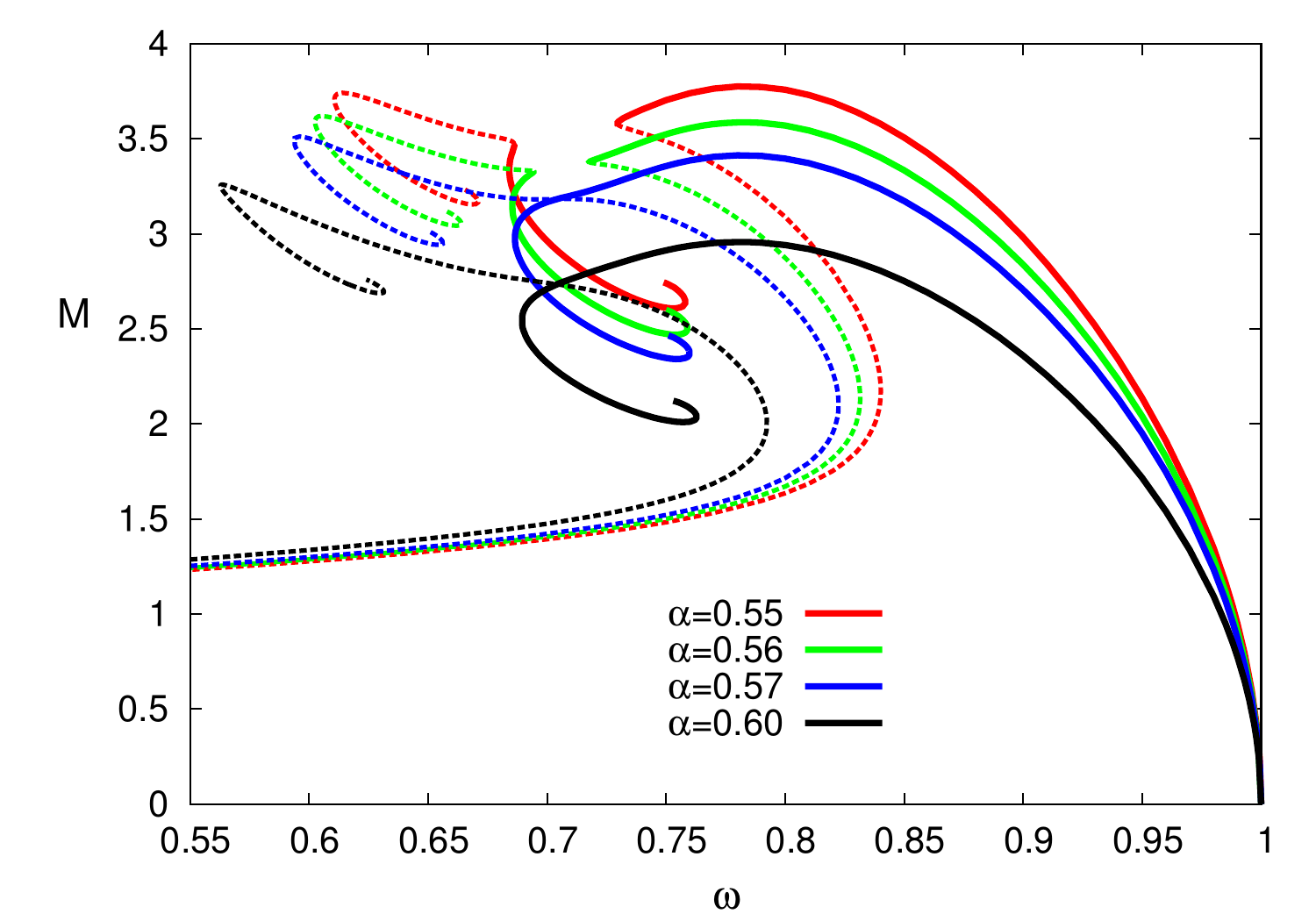}
\\[4pt]
\includegraphics[height=.31\textheight,  angle =-90]{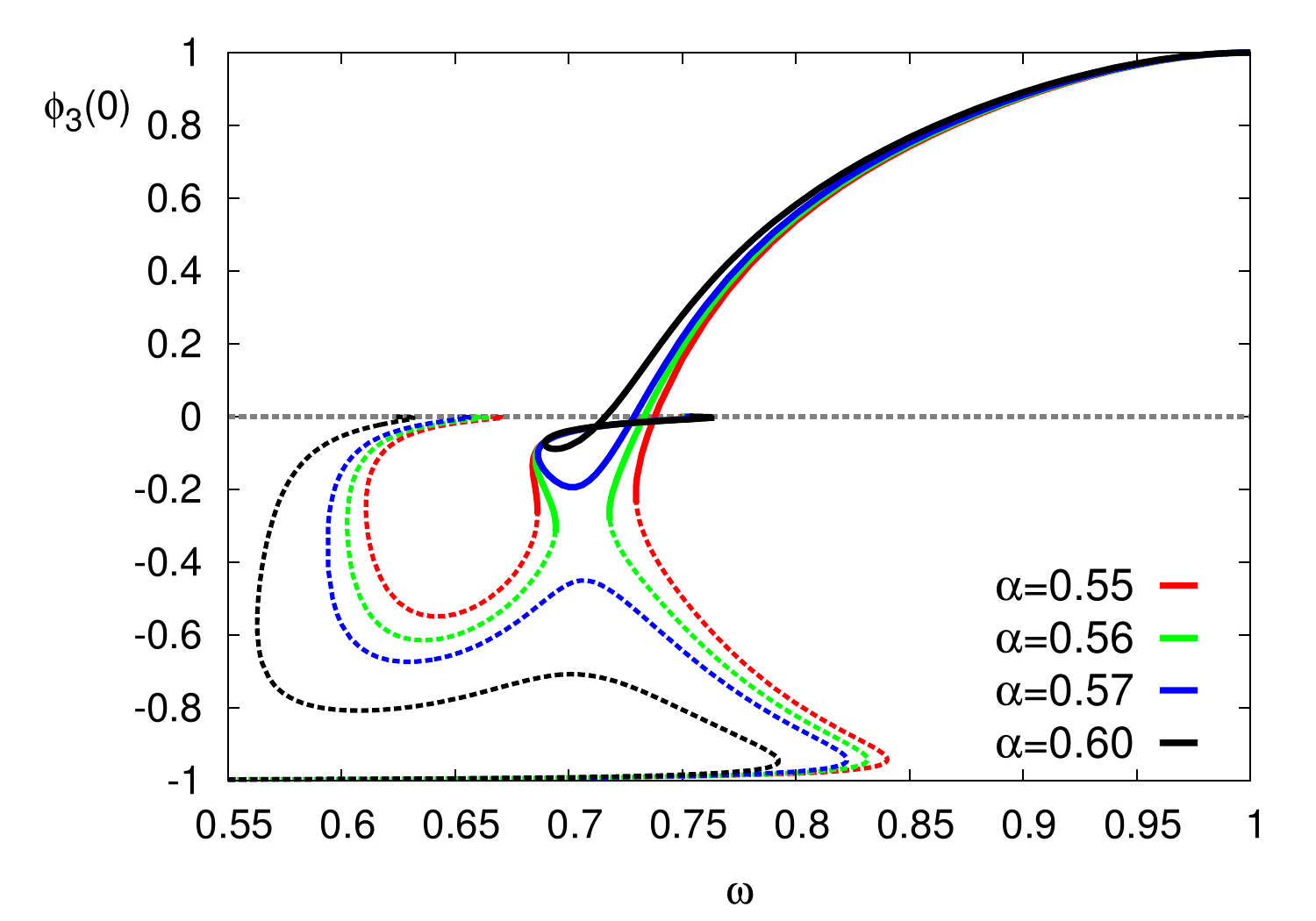}
\includegraphics[height=.31\textheight,  angle =-90]{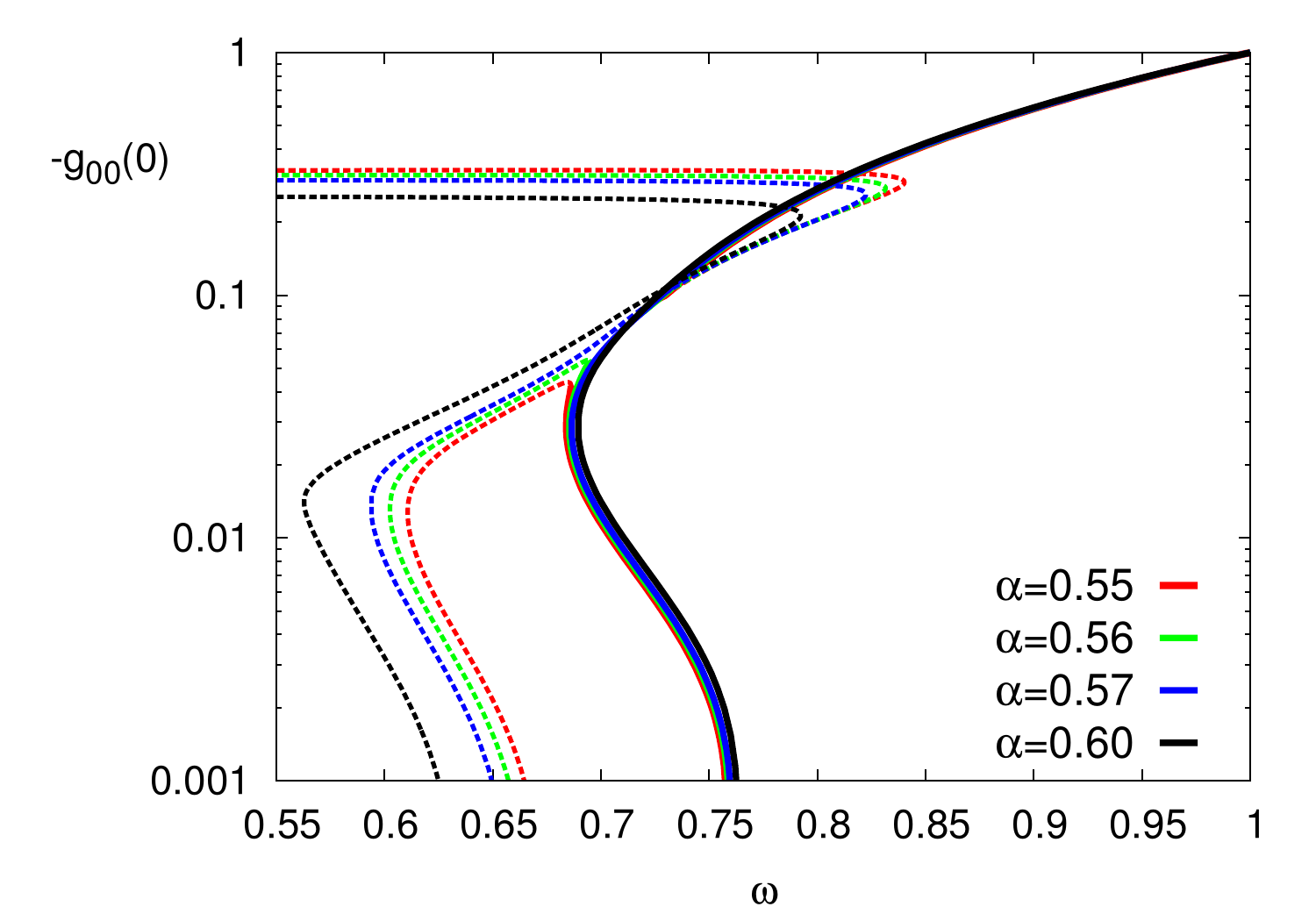}
\end{center}
\caption{\small Bifurcation of two types of spherically symmetric ungauged ($e=0$) $O(3)$ boson stars: 
The ADM mass $M$ in units of $8\pi$ (upper plot), the central values of the scalar component $\phi_3(0)$,(bottom left) and the metric component $-g_{00}(0)$ (bottom right) vs the frequency $\omega$ are plotted for some set of values of the gravitational coupling $\alpha$ and $\beta=0.5$. 
The solid and dashed lines correspond to solutions of types I and II, respectively.}    
\lbfig{fig4}
\end{figure}

We begin by considering the properties of the ungauged spherically symmetric self-gravitating localized solutions of the model \re{lag}, fixing at first the rescaled parameter $\beta=0.5$ and subsequently considering the variation of it. 
A peculiar feature of the model is that, depending on the coupling constants, it can support two different types of solitonic solutions. 

The solutions of the first type are similar to the usual mini-boson stars with global $U(1)$ symmetry in the Einstein-Klein-Gordon theory \cite{Liebling:2012fv}, and to the solitonic solutions of the $O(3)$-sigma model with the simple pion mass potential \cite{Herdeiro:2018djx,Mikhaliuk:2025mxy}, which do not possess a flat space limit. 
These type I boson stars arise smoothly as perturbations around the vacuum $\phi^a =(0,0,1)$ in Minkowski spacetime, when the angular frequency $\omega$ is decreased below its maximal value $\omega_{max}^{I} = \mu =1$ (in rescaled quantities).
This is often referred to as the Newtonian limit.

When decreasing $\omega$ below the mass threshold the ADM mass $M$ of the type I boson stars increase, until a maximum is reached, as seen in Figs.~\ref{fig1} and \ref{fig2}, upper plots. 
The pattern of the subsequent evolution depends on the value of the effective gravitational constant $\alpha$. 
As $\alpha > \alpha_{cr} \approx  0.56$, the $O(3)$ boson stars exhibit the typical spiraling and oscillating pattern with successive backbendings of the usual boson stars until reaching a singular limiting solution. 
Thus the type I boson stars exist within a restricted interval of values of the angular frequency $\omega \in [\omega_{min}^{(I)}, 1]$.
At the minimal value of the frequency $\omega_{min}^{(I)}$ a forward branch arises, starting the formation of the spiral.

Both the mass and the charge of the $O(3)$ boson stars tend to finite limiting values at the centers of the spirals. 
When following the spirals towards their centers, the $O(3)$ field $\phi^a$ rotates from the vacuum $\phi^a =(0,0,1)$ toward its central critical solution with $\phi^a_{cc}(0) =(\phi_{cc},\sqrt{1-\phi_{cc}^2},0)$, as displayed in Figs.~\ref{fig1} and \ref{fig2}, bottom left plots, for $\phi^3(0)$. 
Similar to the case of the usual boson stars, the metric function $g_{00}$ at the center of the configuration approaches zero in this limit, as seen in Fig.~\ref{fig1}, bottom right plot. 
When following the spiral, the core of the configuration shrinks.
This suggests that the profile of the central critical solution might correspond to a very thin wall separating a compact domain $\phi^3 =0$ and the vacuum $\phi^3 =1$.  
However, a different numerical approach will be necessary to study this limit.

As seen in Fig.~\ref{fig1}, there is a second type of boson stars, residing in the frequency range $0 = \omega_{min}^{II} \le \omega \le \omega_{max}^{II}$. 
These arise from the flat space static $(\omega=0)$ non-topological solitons of the $O(3)$ sigma model with potential \re{pot-mod}  \cite{Ferreira:2025xey,Verbin:2007fa}, when coupled to gravity.
These solutions owe their existence to the partly negative self-interaction potential \cite{Nucamendi:1995ex,Bechmann:1995sa,Kleihaus:2013tba,Chew:2024bec}.
The distinctive new feature of such solutions is that, as $\omega$ decreases toward zero, the central value of the scalar component $\phi^3(0)$ tends to its value at the \textsl{anti-vacuum}, where the scalar triplet assumes the values $(0,0,-1)$, corresponding to the negative minimum of the potential.
This can be inferred from Fig.~\ref{fig3}, bottom left plot, for the configuration I. 
Notably, limiting solutions at $\omega=0$ possess finite mass while their Noether charge is vanishing. This limit corresponds to the self-gravitating non-topological soliton of the non-linear $O(3)$ sigma model, whose properties in flat space were discussed in \cite{Ferreira:2025xey}.

These type II boson stars mimic to some extent the qualitative pattern of the dynamical evolution of the behavior of the type I boson stars.
Thus they exhibit the typical spiraling pattern seen also in the evolution of the type I solutions toward a singular limiting configuration. 
Within the spiral, the central value of the $O(3)$ field $\phi^a$ rotates from the anti-vacuum  $\phi^a =(0,0,-1)$ toward a second central critical solution with $\phi^a_{cc}(0) =(\phi_{cc},\sqrt{1-\phi_{cc}^2},0)$, as displayed in Figs.~\ref{fig1} and \ref{fig3}, bottom left plots. 
Again, the central value of the metric function $g_{00}$ approaches zero in this limit. 
We note that the size of the spiral can be very small. 
Here, a tiny change of $\omega$ can strongly affect the force balance between the gravitational attraction and the scalar interaction, as illustrated in Fig.~\ref{fig1}, upper left zoomed subplot. 
We also note that, for some maximal value of $\alpha$ the type II solutions then cease to exist.

\begin{figure}[t!]
\begin{center}
\includegraphics[height=.38\textheight,  angle =-0]{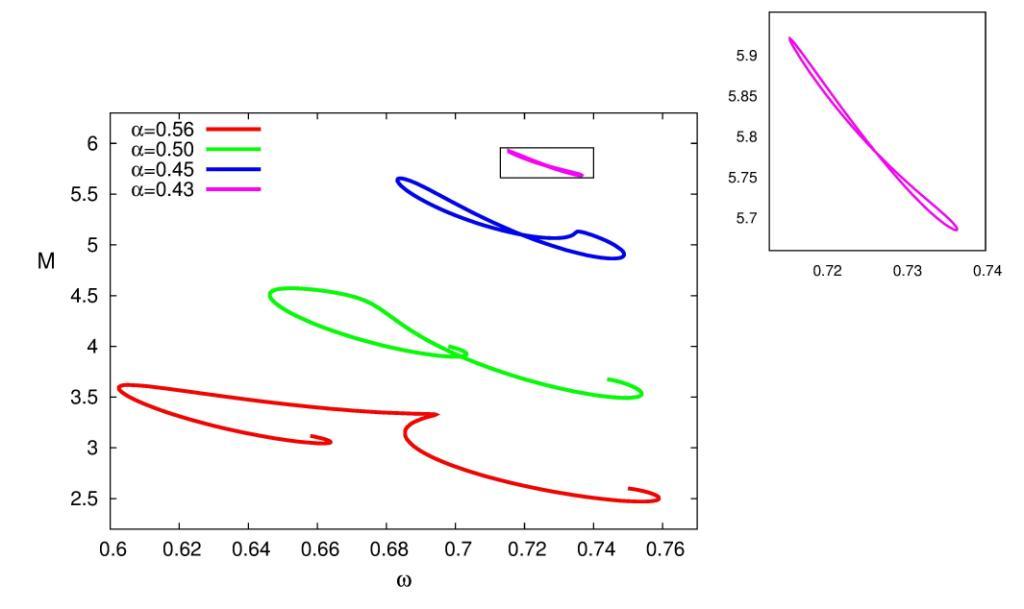}
\\[4pt]
\includegraphics[height=.31\textheight,  angle =-90]{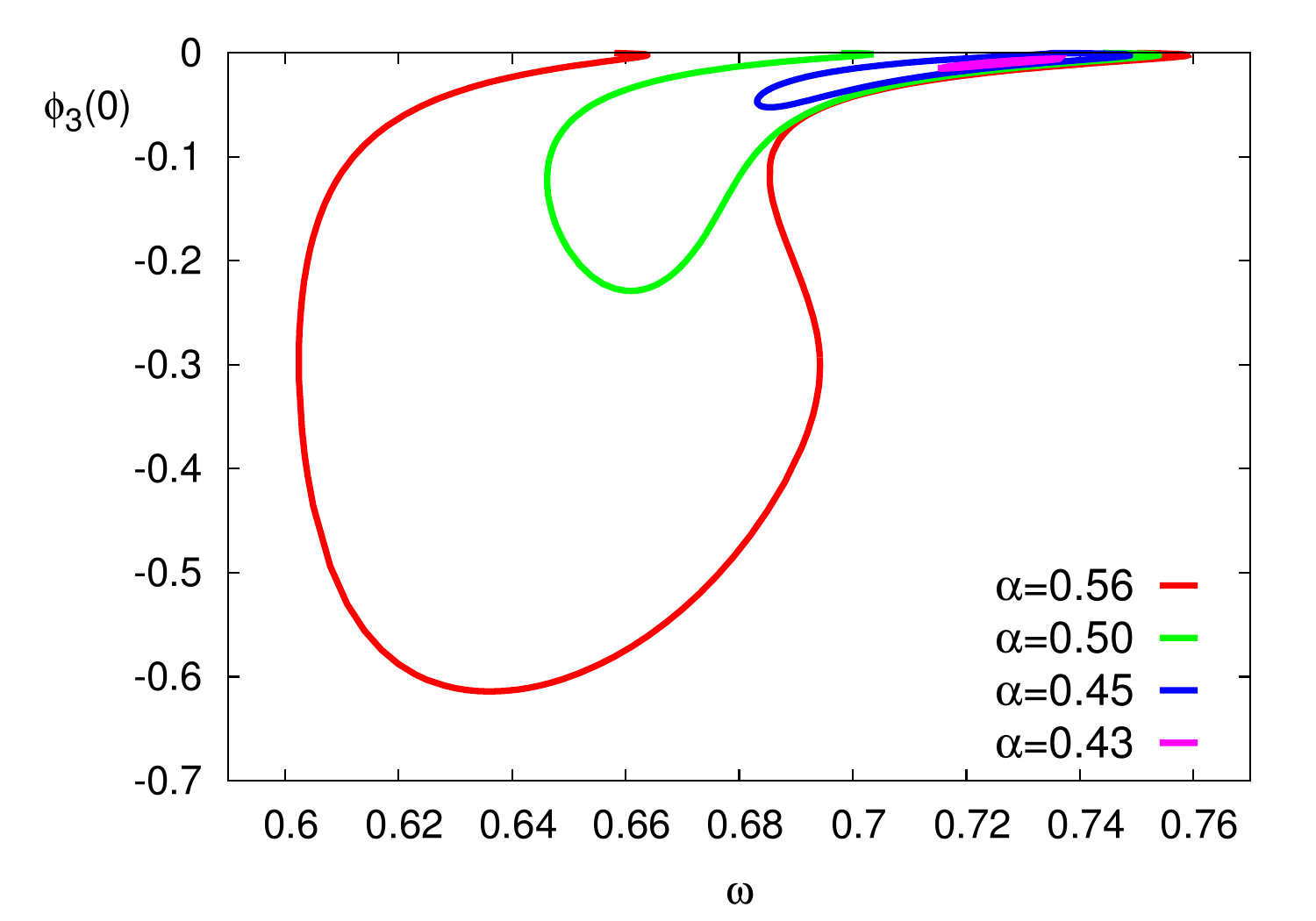}
\includegraphics[height=.31\textheight,  angle =-90]{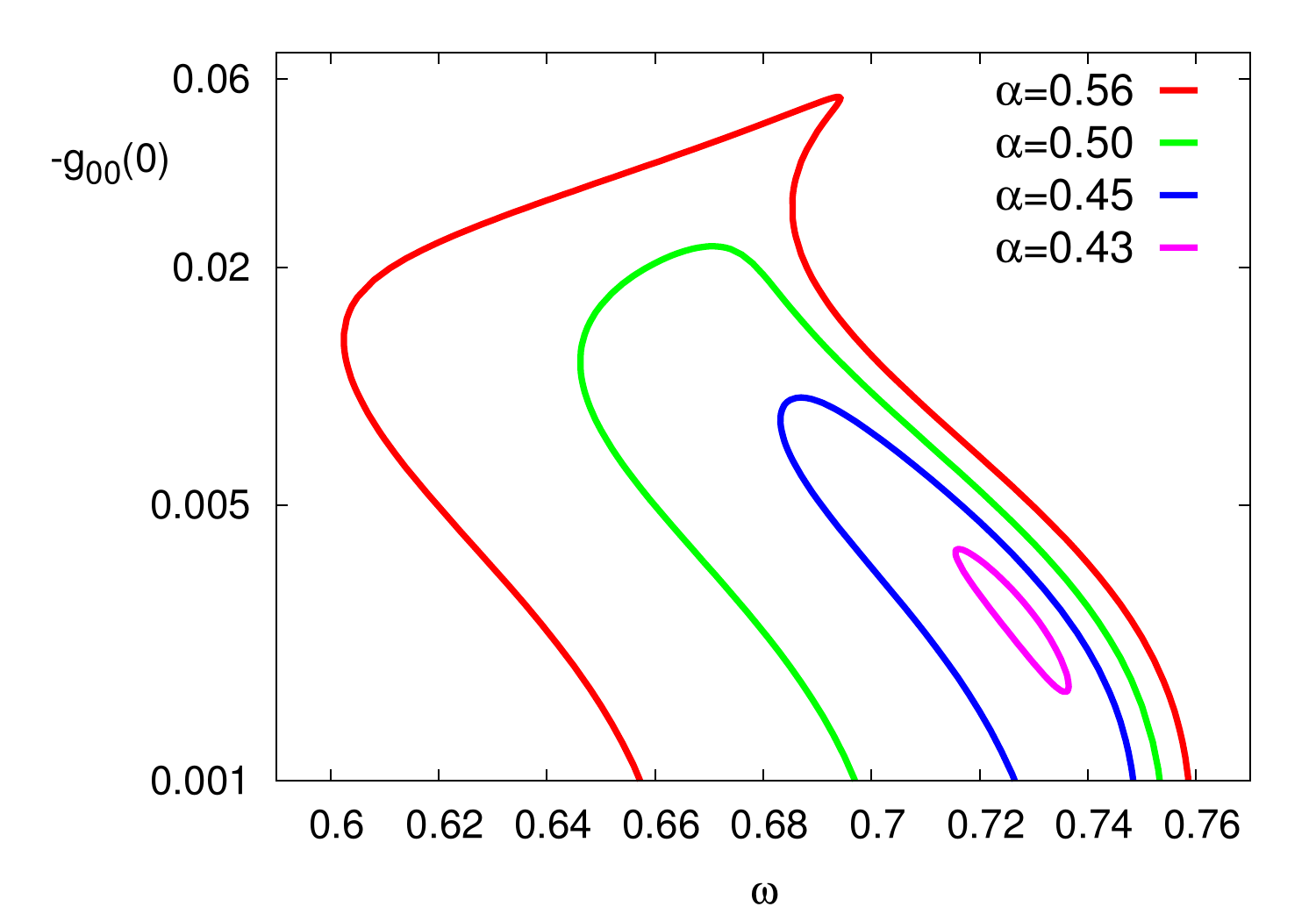}
\end{center}
\caption{\small Evolution of the detached double spirals of ungauged ($e=0$) $O(3)$ boson stars in an isolated domain: 
The ADM mass $M$ in units of $8\pi$ (upper plot), the central values of the scalar component $\phi_3(0)$,(bottom left) and the metric component $-g_{00}(0)$ (bottom right) vs the frequency $\omega$ are plotted for some set of values of the gravitational coupling $\alpha$ and $\beta=0.5$.}    
\lbfig{fig5}
\end{figure}

\begin{figure}[t!]
\begin{center}
\includegraphics[height=.38\textheight,  angle =-90]{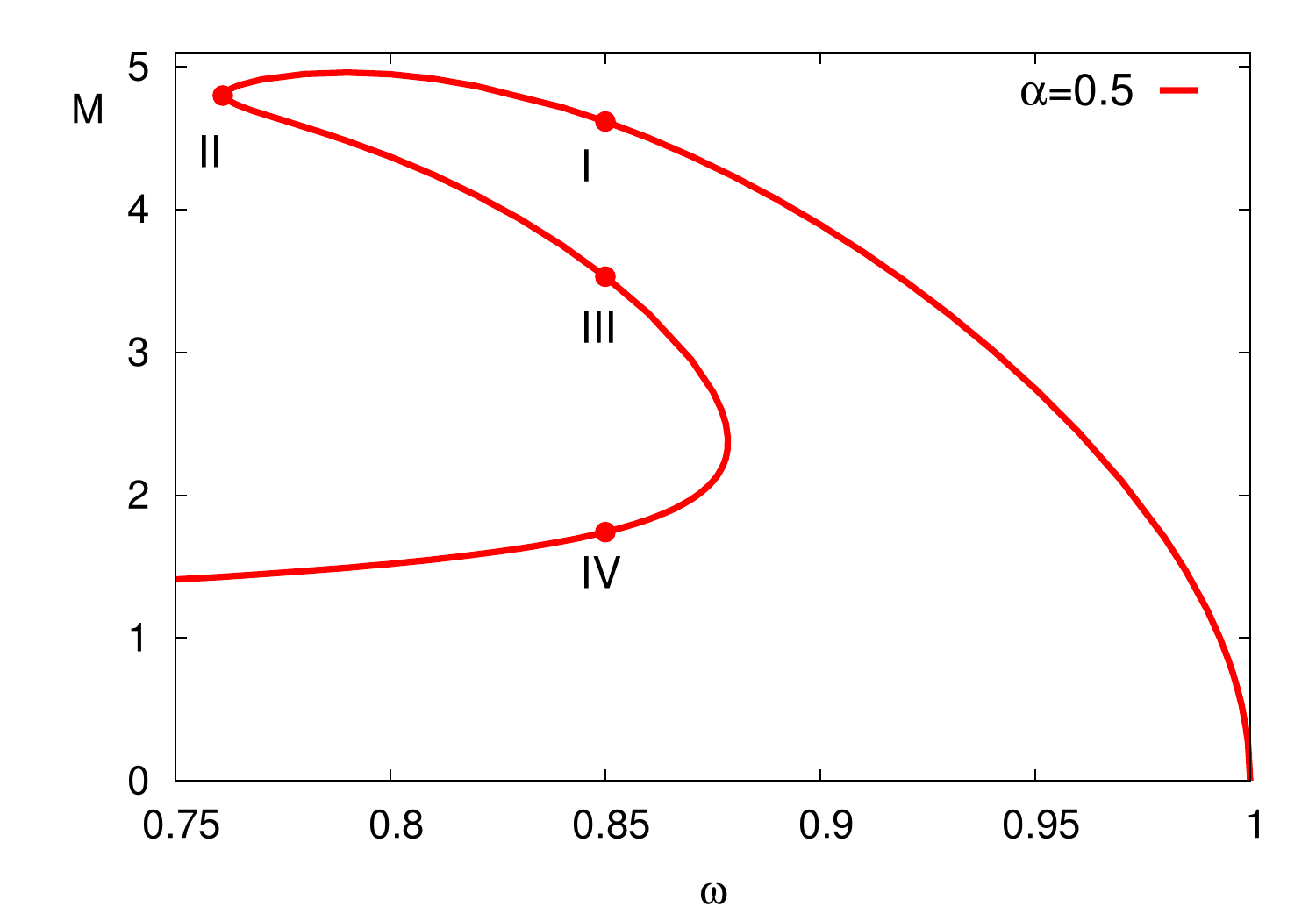}\\
\includegraphics[height=.32\textheight,  angle =-90]{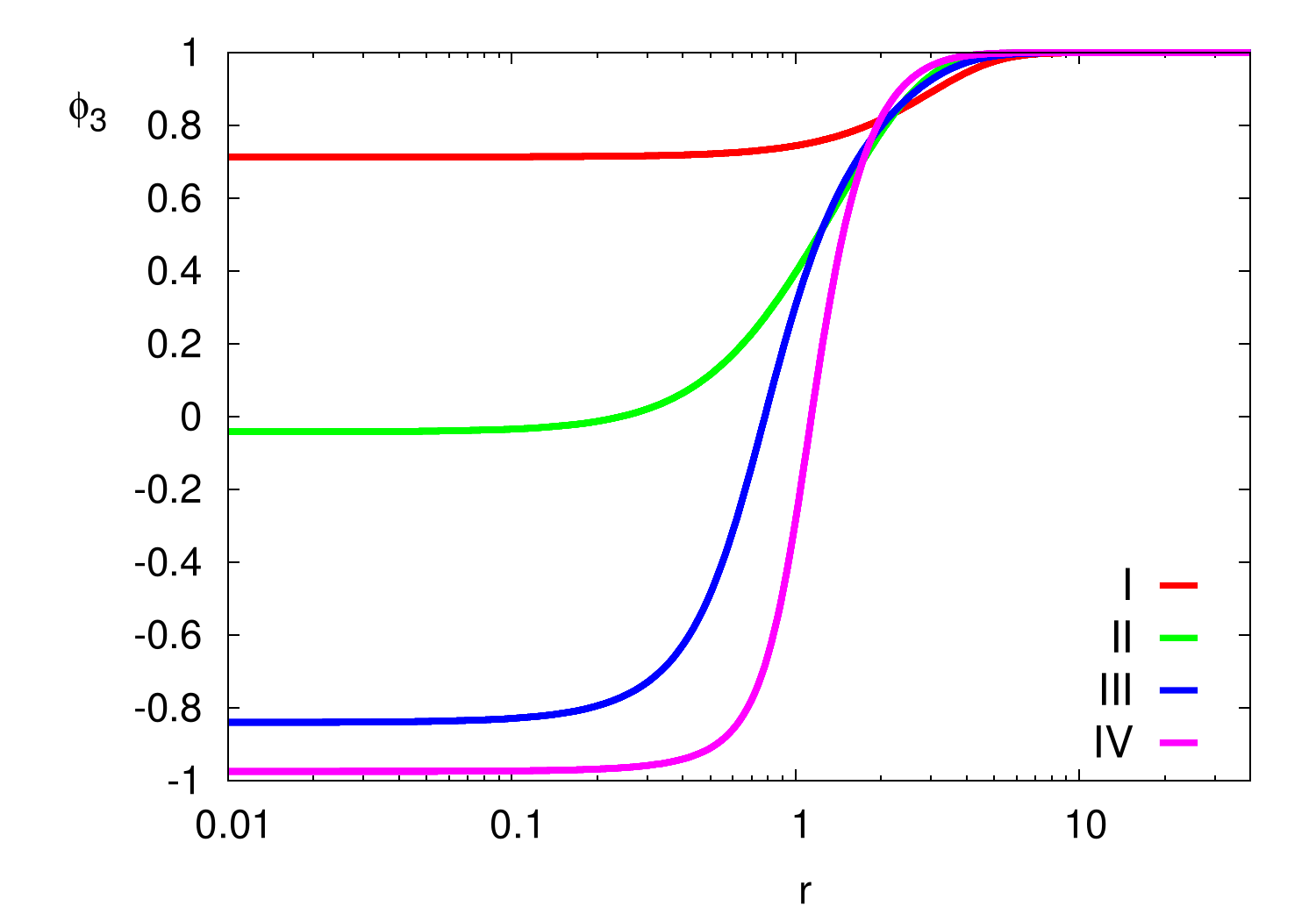}
\includegraphics[height=.32\textheight,  angle =-90]{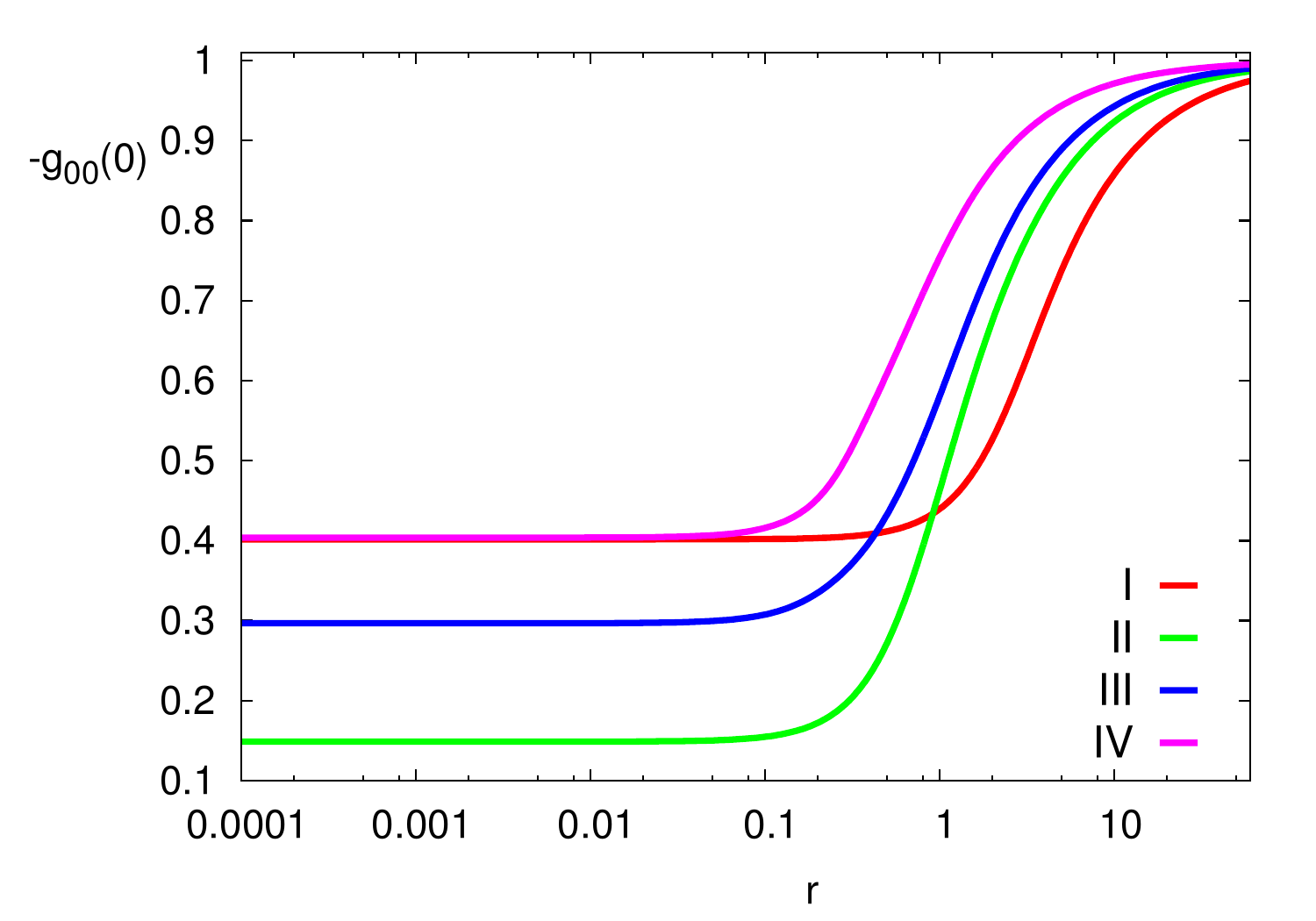}
\end{center}
\caption{\small Spherically symmetric ungauged ($e=0$) $O(3)$ boson stars: 
The ADM mass $M$ of the merged sets of type I and type II boson stars in units of $8\pi$ vs the frequency $\omega$ for $\alpha=0.5$ and $\beta=0.5$ (upper plot) and illustrative radial profiles of the scalar component $\phi_3$ (bottom left) and the metric component $-g_{00}$ (bottom right) of the solutions, indicated by dots on the upper plot.}    
\lbfig{fig6}
\end{figure}

The above picture changes dramatically as $\alpha \to \alpha_{cr}$.
While for larger values of $\alpha$ the type II boson stars are disconnected from the gravitating solitons of the first type, as seen in the mass-frequency diagram in Fig.~\ref{fig1}, upper plot, both types start to overlap with decreasing $\alpha$.
Thus $\omega_{max}^{(II)}<\omega_{min}^{(I)}$ changes to $\omega_{max}^{(II)}>\omega_{min}^{(I)}$, leading to an intersection of both types of solutions in the mass (charge)-frequency domain.
Focusing on this intersection, we note that it develops into a bifurcation of both types at a critical value of the gravitational coupling $\alpha_{cr}$, as illustrated in Fig.~\ref{fig4}.

Indeed, here we see for $\alpha> \alpha_{cr}$ the first branch of the type I solutions and the second branch of the type II solutions intersect in the $M$-$\omega$ diagram.
These branches then also intersect in the $g_{00}$-$\omega$ diagram at a similar value of the frequency.
Moreover, we note in the $\phi_3$-$\omega$ diagram that the minimum of the type I solutions and the local maximum of the type II solutions approach each other with decreasing $\alpha$.
Finally, as $\alpha = \alpha_{cr}$, the solutions of both types bifurcate at a critical value of the frequency. 
Subsequently, the spirals become detached from their previous branches and reconnect among themselves, forming a double spiral structure.
At the same time, the remaining branches of both types of solutions reconnect with each other, forming a connected set of solitonic solutions of the $O(3)$ sigma model, which exists for the whole range of values of the frequency $\omega \in [0,1]$.

Focusing further on the evolution of the detached double spirals in Fig.~\ref{fig5}, we note their metamorphosis to a closed loop that intersects itself, as $\alpha$ is decreased further.
This pattern is also highlighted in Fig.~\ref{fig1}, right upper subplot.
With further decrease of $\alpha$, these remnants of the detached double spirals disappear.
On the other hand, we illustrate the evolution of the solutions along the remaining connected set of solutions that span the full frequency range $0 \le \omega \le 1$ further in Fig.~\ref{fig6}, where we exhibit several configurations for the connected set of type I and type II solutions at $\alpha=0.5$.
Moreover, Fig.~\ref{fig1} shows that for small $\alpha$ the maximum of the curves increases strongly with decreasing $\alpha$.
While this might suggest a divergence of the mass in the flat space limit for $\omega \to \omega_{max}$, as known from ordinary Q-balls, further calculations with another numerical method will be needed to clarify the limit.

\begin{figure}[t!]
\begin{center}
\includegraphics[height=.285\textheight,  angle =0]{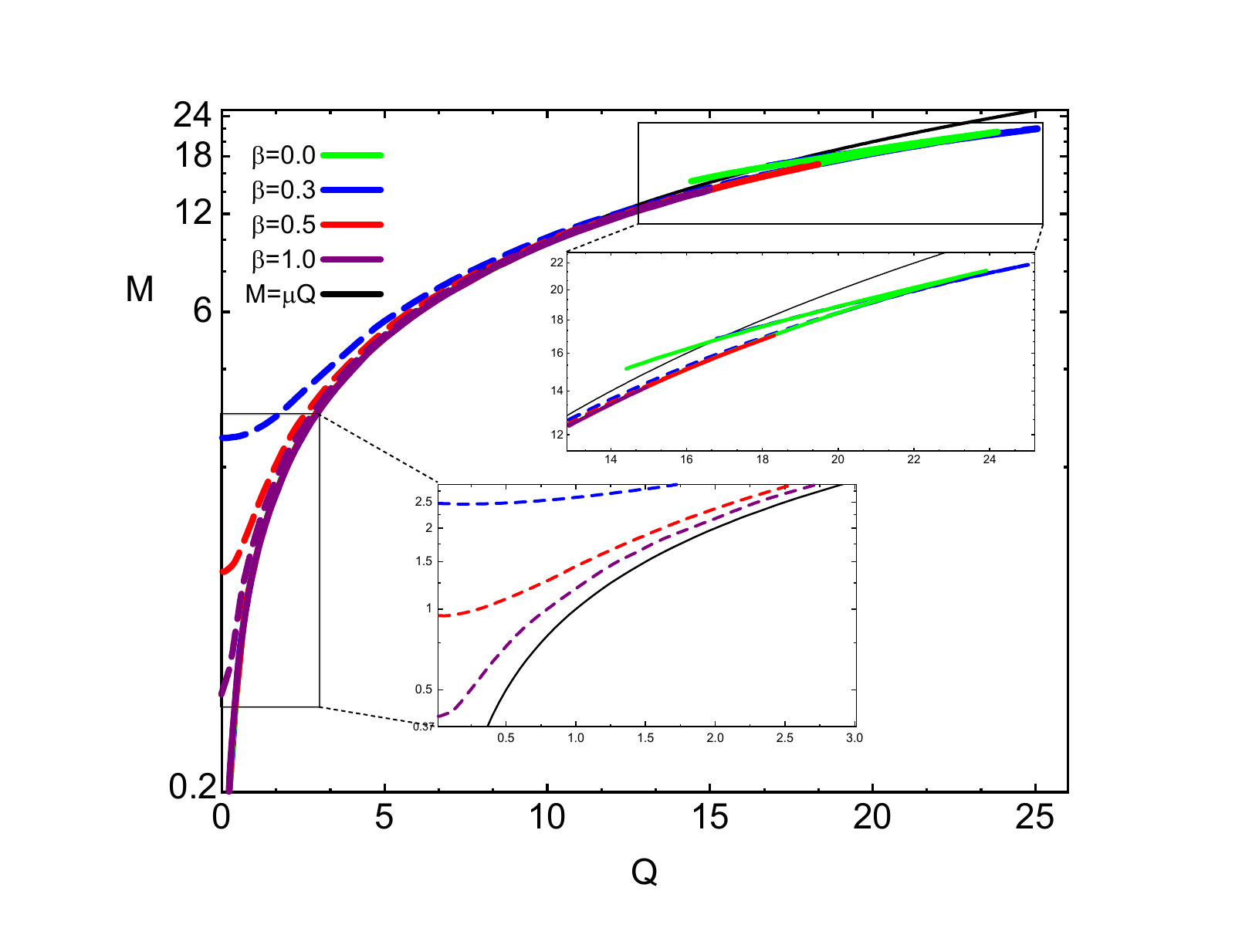}
\includegraphics[height=.278\textheight,  angle =0]{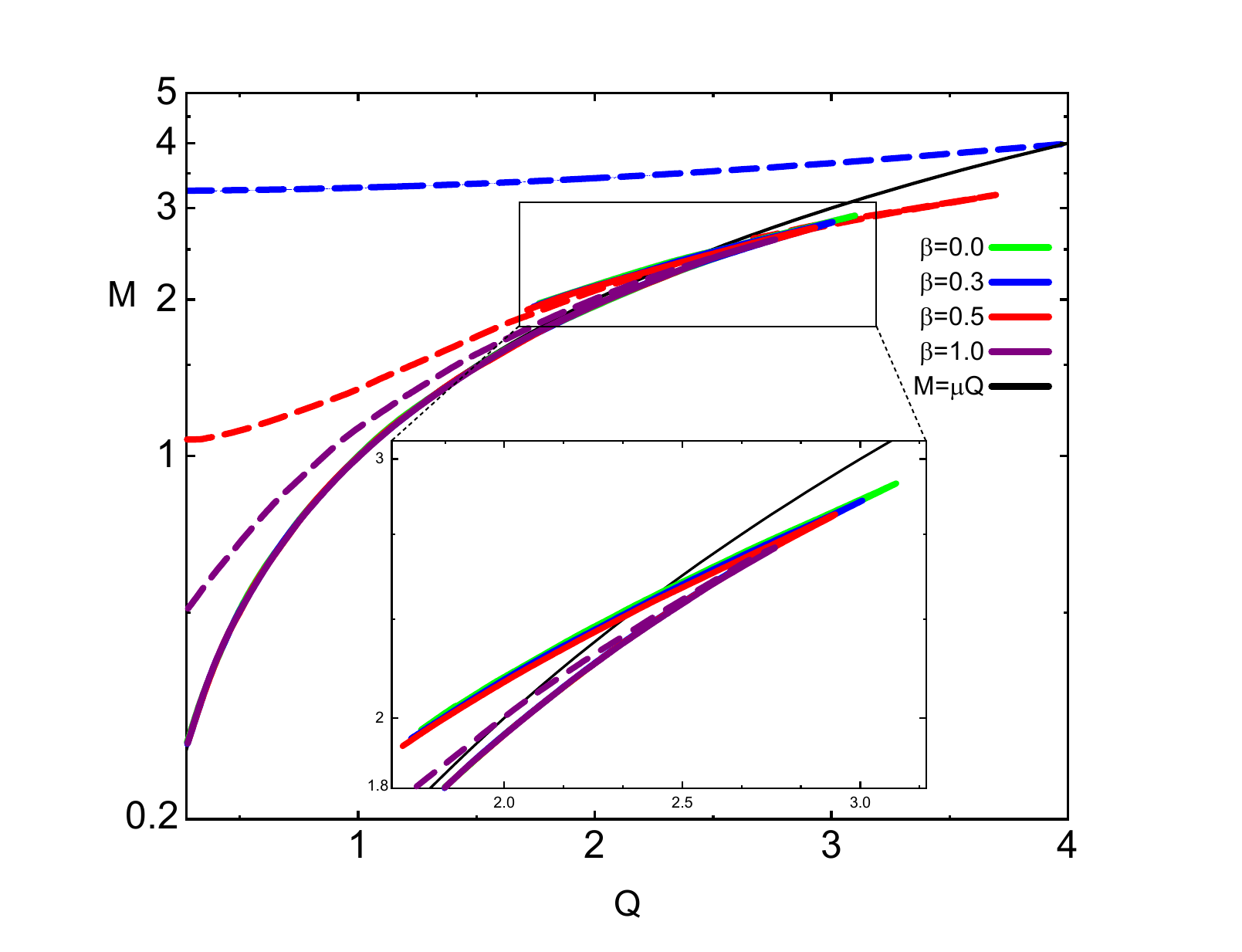}
\end{center}
\caption{\small Spherically symmetric ungauged $(e=0)$ $O(3)$ boson stars: 
The ADM mass $M$ of type I and type II boson stars vs the Noether charge $Q$, both in units of $8\pi$, for $\alpha=0.3$ (left plot) and for $\alpha=0.6$ (right plot) and some set of values of $\beta$. 
The solid and dashed lines correspond to solutions of types I and II, respectively. 
The curve $M=\mu Q$ (black solid line) indicates the mass of Q free scalar quanta.}  
\lbfig{fig7}
\end{figure}

\begin{figure}[h!]
\begin{center}
\mbox{ \hspace*{-1.0cm}
\includegraphics[height=.32\textheight,  angle =-90]{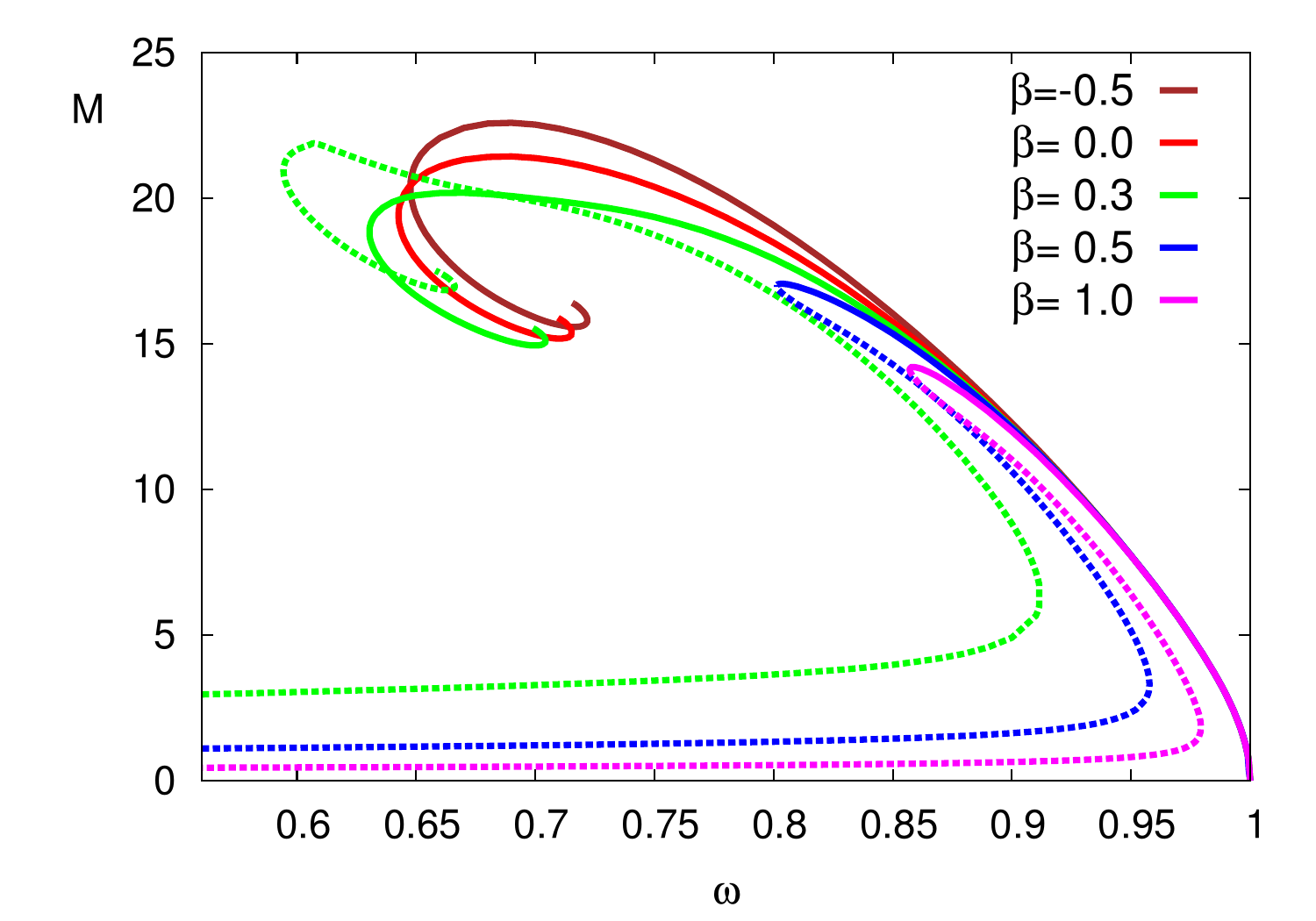}
\includegraphics[height=.32\textheight,  angle =-90]{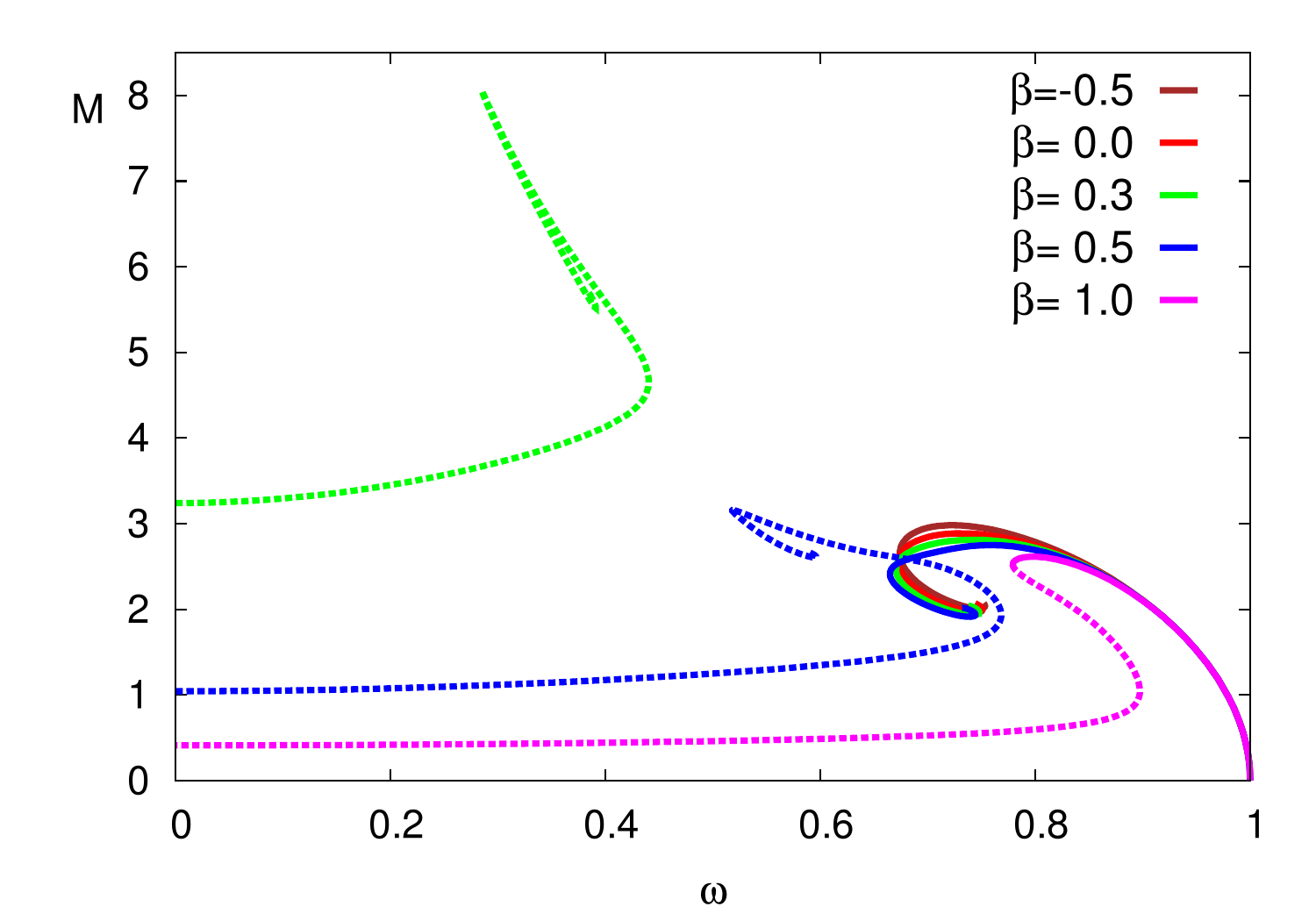}
}
\end{center}
\caption{\small Spherically symmetric ungauged ($e=0$) $O(3)$ boson stars: 
The ADM mass $M$ in units of $8\pi$  vs the frequency $\omega$ is plotted for values of the gravitational coupling $\alpha=0.3$ (left plot) and $\alpha=0.6$ (right plot) for some set of values of the rescaled parameter $\beta$.
The solid and dashed lines correspond to solutions of types I and II, respectively.} 
\lbfig{fig8}
\end{figure}

It is also of interest to analyze the dependence of the solutions on the value of the rescaled parameter $\beta$. 
The limiting case $\beta =0$ corresponds to the type I $O(3)$ boson stars previously discussed in \cite{Herdeiro:2018djx}.
The type II boson stars arise from the flat space ($\alpha=0$) non-topological $O(3)$ solitons discussed in \cite{Ferreira:2025xey}. 
They exist for $\beta > \frac{1}{8}$, where the potential \re{pot-mod} possesses a global minimum at $f = \pi$, where $U(\phi) < 0$. 
Thus, the minimal value of the potential is negative, this condition is necessary for soliton solutions to exist.

The parameters $\alpha$ and $\beta$ control the strength of the interactions: increasing $\alpha$ leads to more attraction, whereas an increase of $\beta$ results in more repulsion.
Variation of the gravitational attraction $\alpha$ and of the potential parameter $\beta$ thus changes the pattern of the solutions, as seen in Figs.~\ref{fig7} and \ref{fig8}.
Figure \ref{fig7} displays the ADM mass $M$ of the solutions versus the Noether charge $Q$ for two values of $\alpha$ and a set of values of $\beta$.
We note that, for a fixed value of $\alpha$, an increase of $\beta$ yields a decrease of the ADM mass of the solutions of both types. 
The mass of the type I boson stars depends only weakly on $\beta$, unlike the mass of the type II solutions, which rapidly increases as $\beta \to 1/8$ from above. 
This pattern is similar to the case of non-topological solitons of the $O(3)$ sigma model with a symmetry-breaking potential in Minkowski space \cite{Ferreira:2025xey}.

Figure \ref{fig8} illustrates the dependence of the bifurcations of the solutions on the parameter $\beta$ for two values of $\alpha$, showing the ADM mass $M$ of the solutions versus the frequency $\omega$.
For $\alpha = 0.3$ (left plot) the bifurcation between the $M(\omega)$ branches of the boson stars of both types occurs at a critical value $\beta \approx 0.38$. 
For $\alpha=0.6$ (right plot) the gravitational attraction is stronger, and the bifurcation occurs at the larger value of $\beta \approx 0.54$. 
We recall that for $\beta=0.5$ such a bifurcation occurs at $\alpha \approx 0.56$.

Interestingly, boson stars of the first type exist also for negative values of the parameter $\beta$, as demonstrated in Fig.~\ref{fig8}. 
Clearly, the properties of the type I boson stars are mainly determined by the gravitational interaction, while for the solutions of the type II the scalar interaction becomes more important.

Concerning the stability of the boson stars, we note that the solutions in Fig.~\ref{fig7} above the black curve $M=\mu Q$ are unbound.
Thus the boson stars above the curve are expected to be unstable. 
In contrast, boson stars on the fundamental branch below the curve are bound.
At each spike of the $M(Q)$ curves, the solutions are expected to acquire an(other) unstable radial mode.

\subsection{$U(1)$ gauged $O(3)$ boson stars}

\begin{figure}[h!]
\begin{center}
\mbox{ \hspace*{-1.0cm}
\includegraphics[height=.32\textheight,  angle =-90]{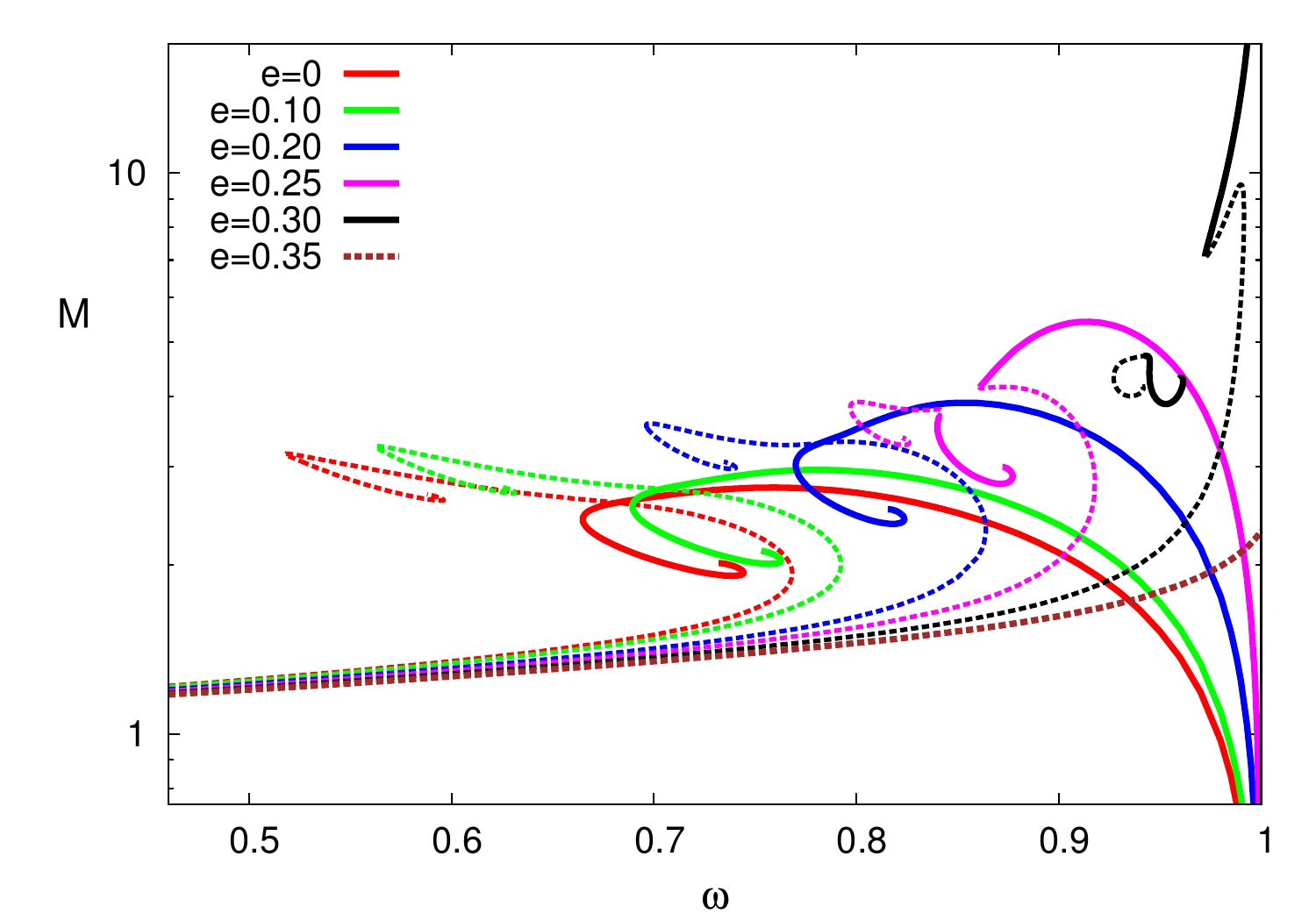}
\includegraphics[height=.32\textheight,  angle =-90]{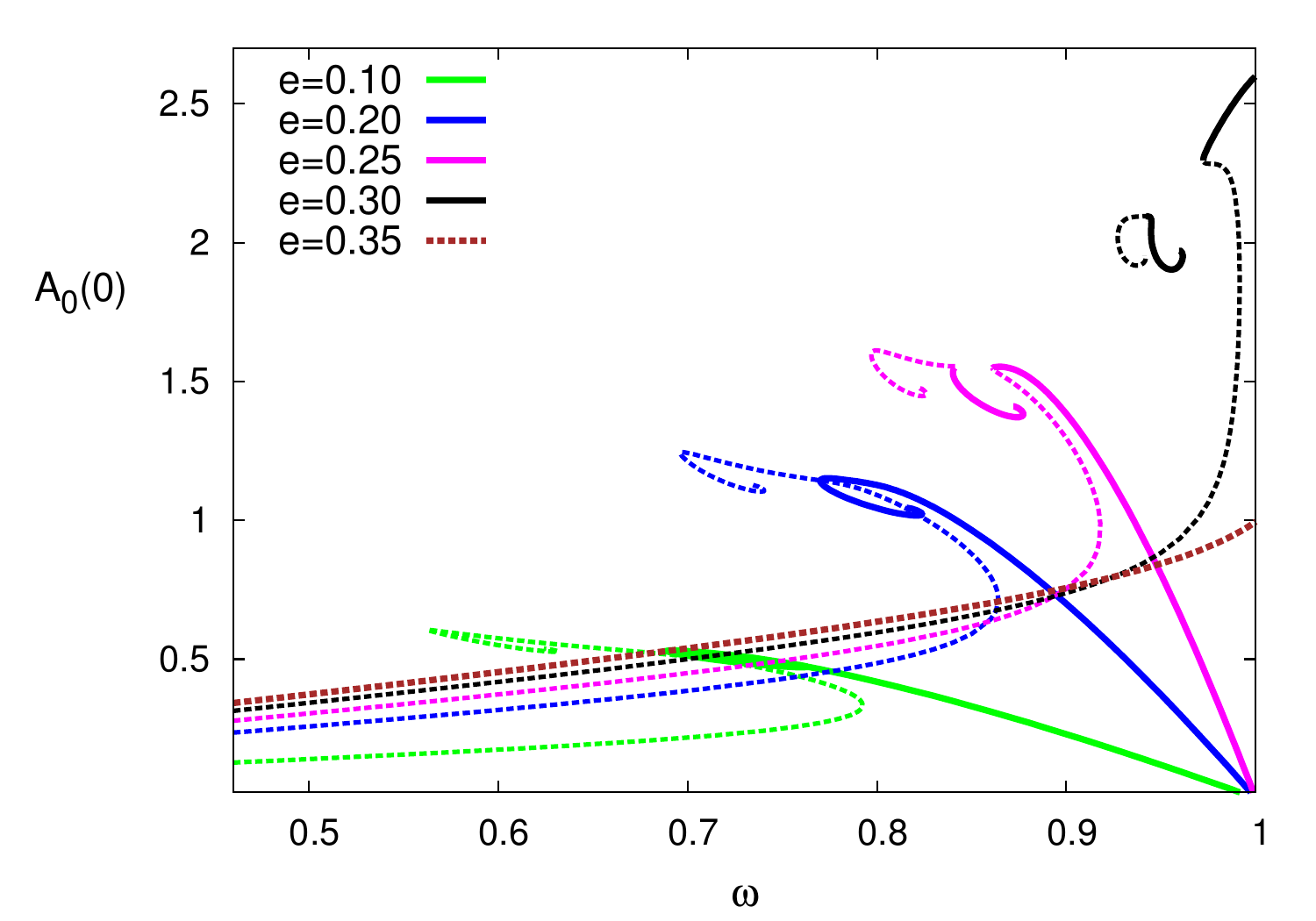}
}
\mbox{ \hspace*{-1.0cm}
\includegraphics[height=.32\textheight,  angle =-90]{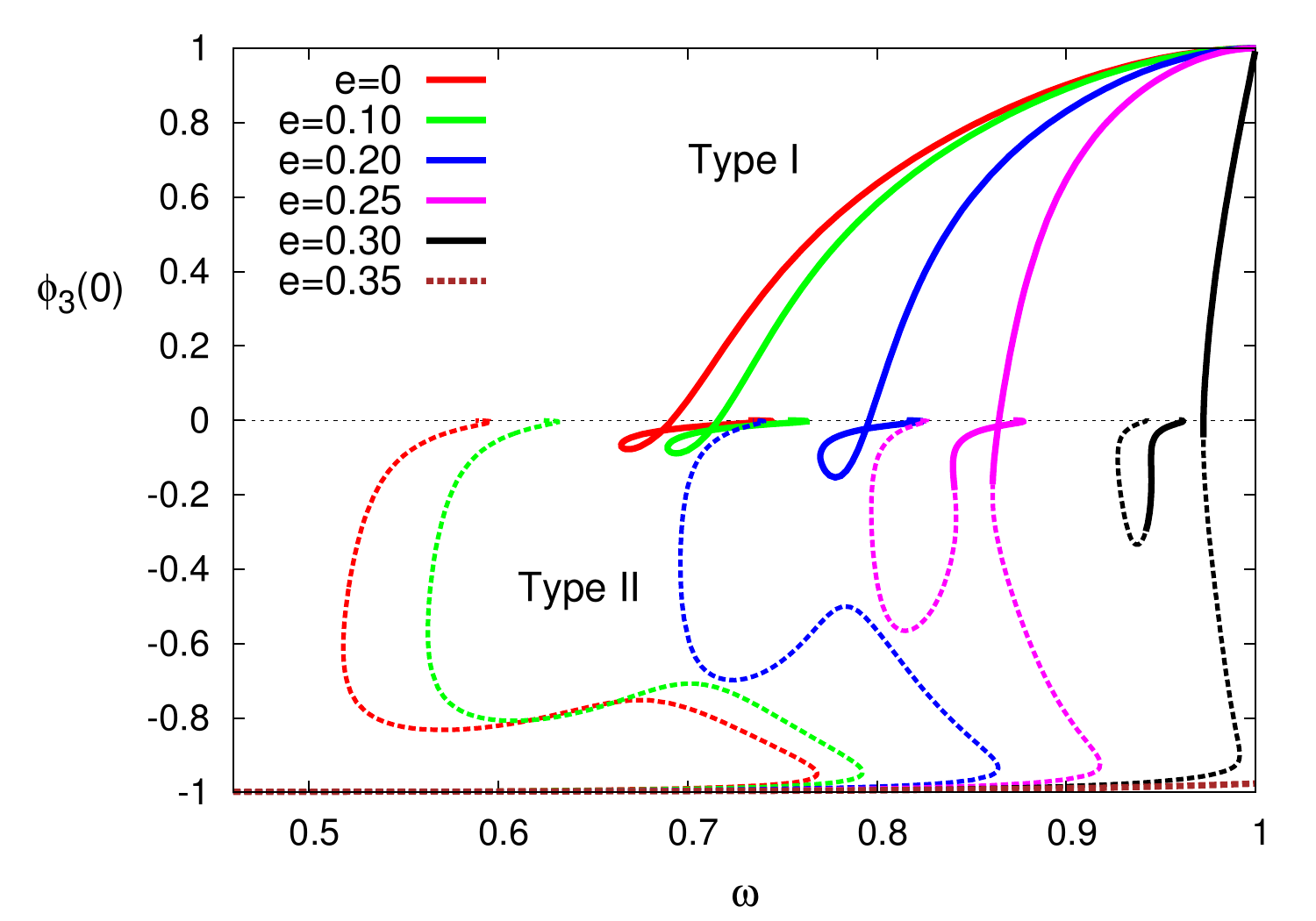}
\includegraphics[height=.32\textheight,  angle =-90]{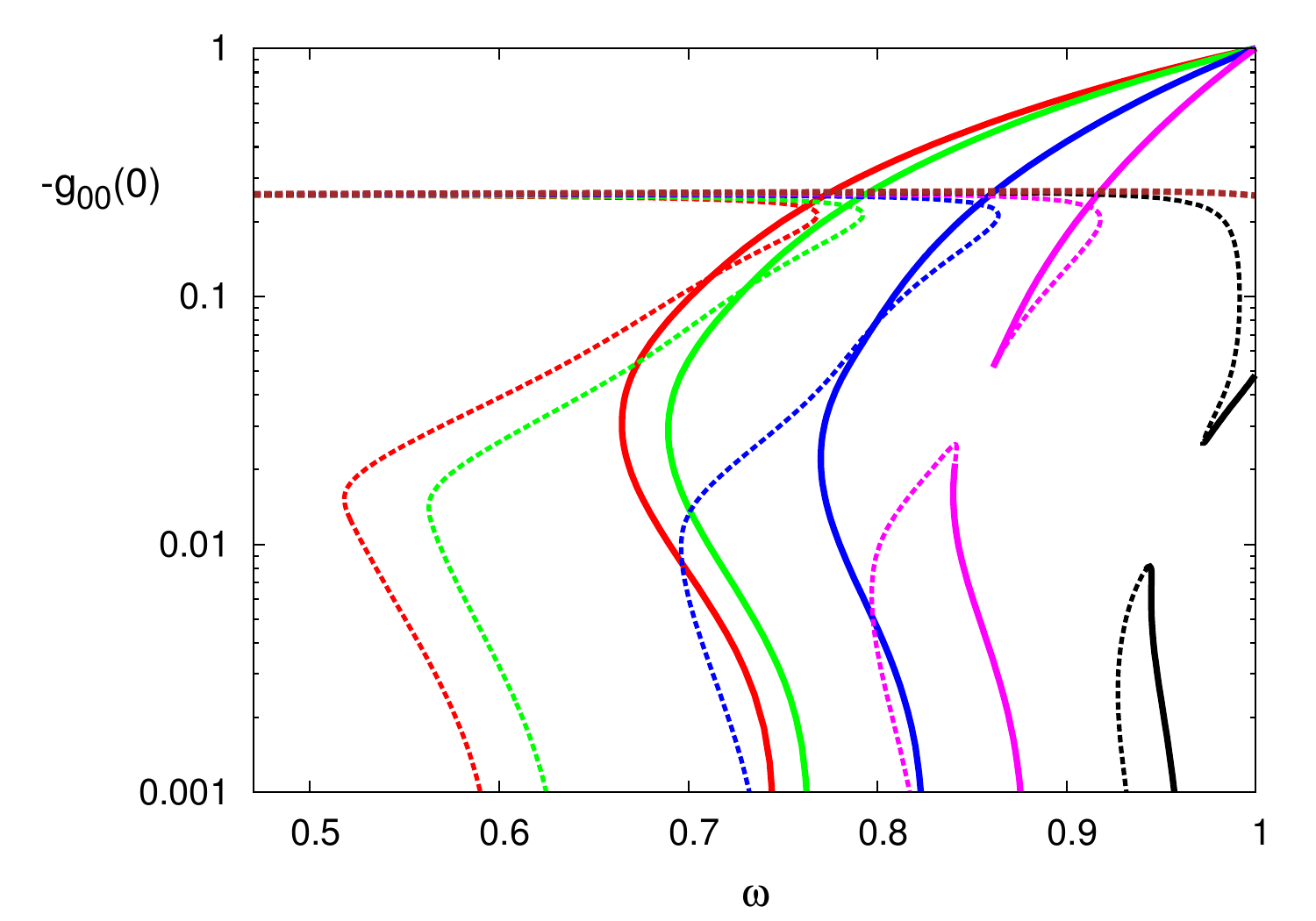}
}
\mbox{ \hspace*{-1.0cm}
\includegraphics[height=.28\textheight,  angle =-0]{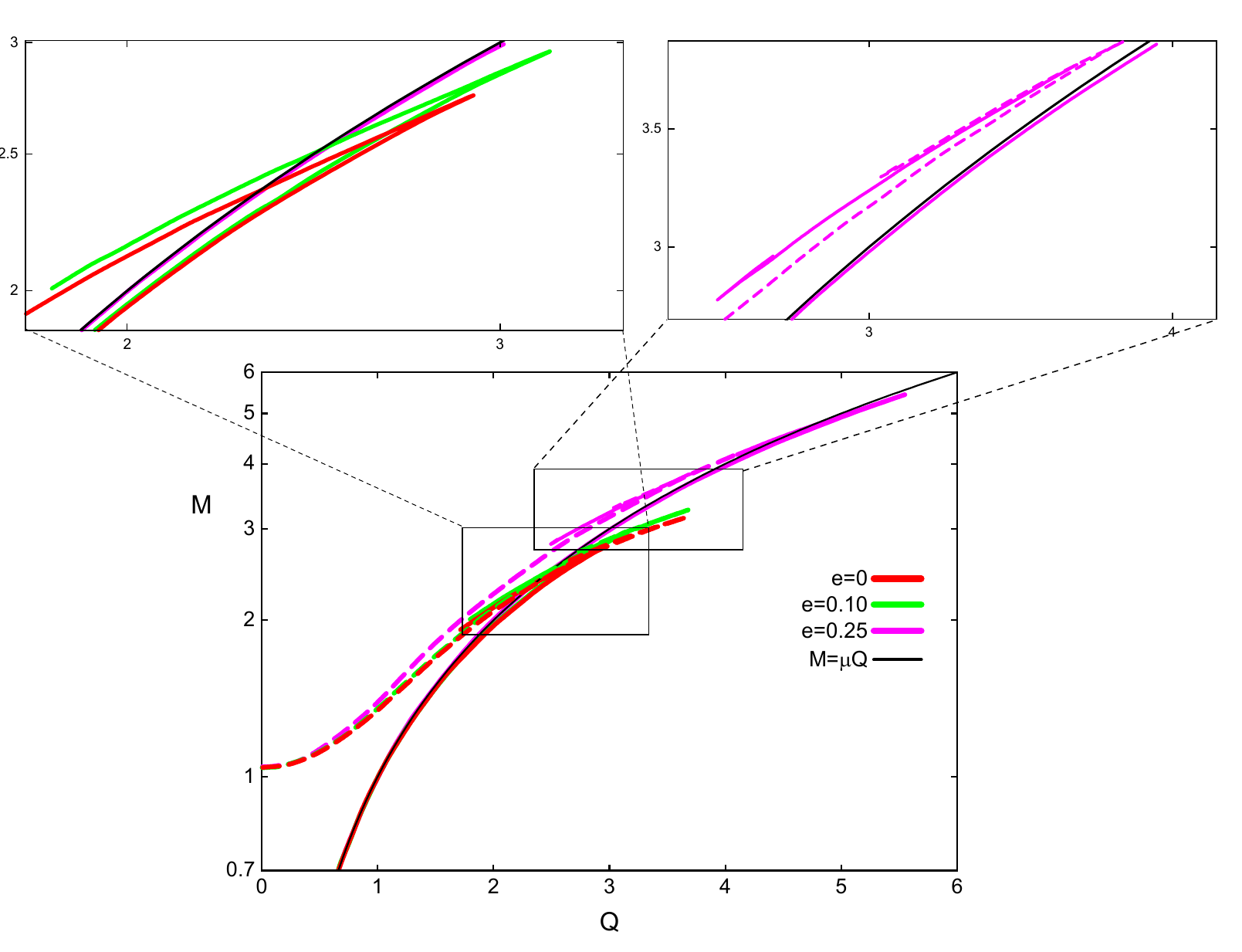}
\includegraphics[height=.28\textheight,  angle =-0]{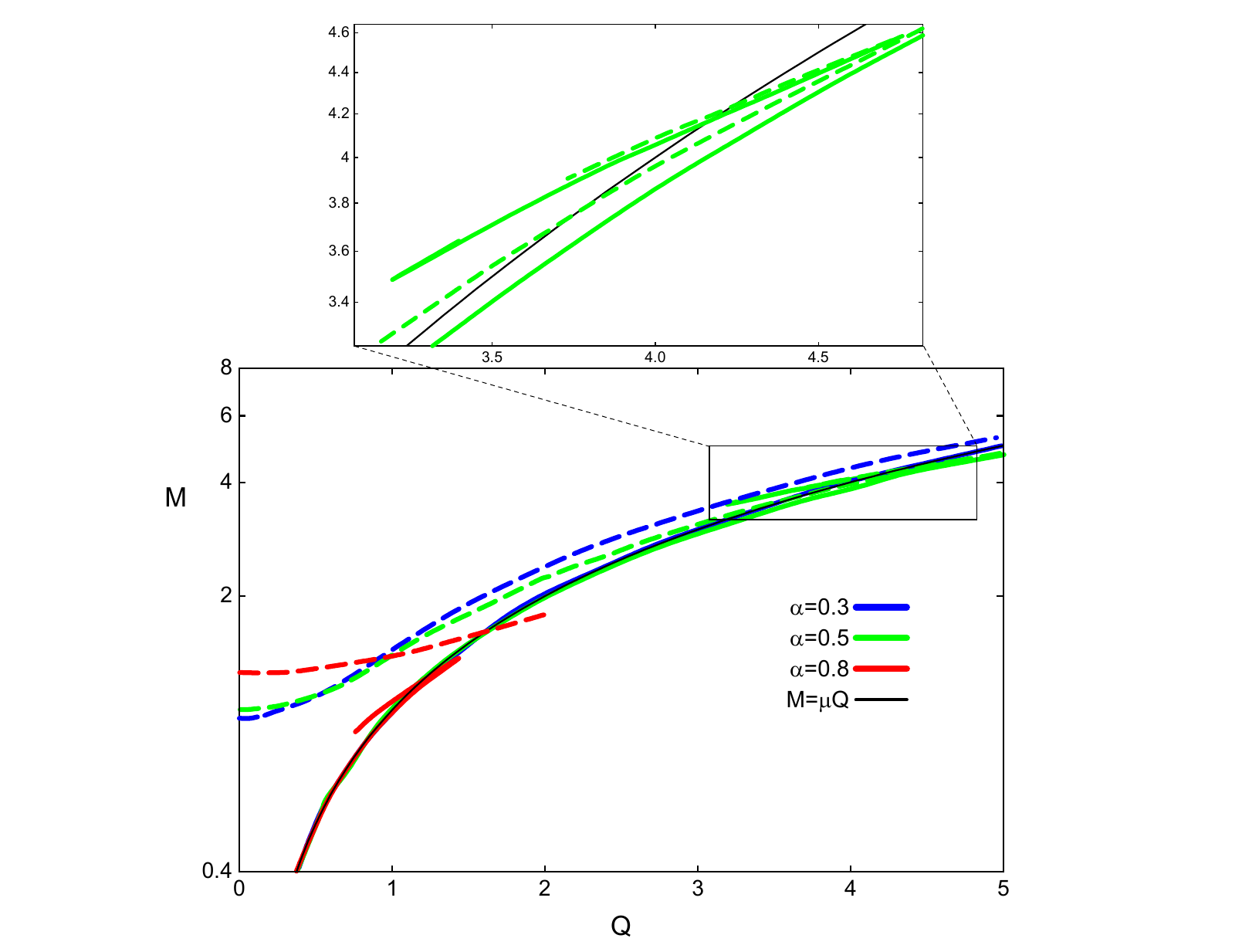}
}
\end{center}
\caption{\small Spherically symmetric $U(1)$ gauged  $O(3)$ boson stars: 
The ADM mass $M$ in units of $8\pi$ (upper left), the central values of the electrostatic potential $A_0$ (upper right), the scalar component $\phi_3$ (middle left), the metric component $-g_{00}$ (middle right)
vs the frequency $\omega$ are plotted for $\alpha=0.6$ and $\beta=0.5$ and some set of values of the gauge coupling $e$.
The ADM mass $M$  vs the charge $Q$, both in units of $8\pi$, is plotted for $\alpha=0.6$ and $\beta=0.5$ and some set of values of the gauge coupling $e$ (bottom left) and for $e=0.1$ and some set of values of the gravitational coupling $\alpha$ and $\beta=0.5$ (bottom right). 
The solid and dashed lines correspond to solutions of types I and II, respectively.  The line $M = \mu Q$ (black) indicates the mass of $Q$ free scalar quanta.}   
\lbfig{fig9}
\end{figure}

We now turn to the $U(1)$ gauged electrically charged boson stars.
Here three parameters, $\alpha,~\beta$ and the gauge coupling $e$, define the relative strength of the interactions in the system. 
An increase in the parameters $\beta$ and $e$ entails an increase of the repulsive force while an increase of the effective gravitational coupling $\alpha$ leads to a stronger attraction.

The $U(1)$ gauged electrically charged boson stars can be constructed by starting from the ungauged solutions considered above and then increasing the value of the gauge coupling constant $e$ (from zero), while keeping the other input parameters fixed. 
The basic properties of the gravitating gauged $O(3)$ boson star solutions so constructed are analogous in many respects to those of the ungauged boson stars of the model.
They are summarized in the following.

First, an increase of the gauge coupling $e$ at fixed $\alpha$ leads to an increase of both the minimal frequency of the type I solutions $\omega_{min}^{(I)}$ and the maximal frequency of the type II solutions $\omega_{max}^{(II)}$, as seen in Fig.~\ref{fig9}.
Moreover, an increase of the gauge coupling $e$ at fixed $\alpha$ leads to an analogous critical behavior of both types of boson star solutions at a critical value of the gauge coupling $e_{cr}$, as seen in the previous subsection at the critical gravitational coupling $\alpha_{cr}$.
Clearly, $e_{cr}$ is a function of the gravitational coupling $\alpha$.

\begin{figure}[t!]
\begin{center}
\includegraphics[height=.32\textheight,  angle =-90]{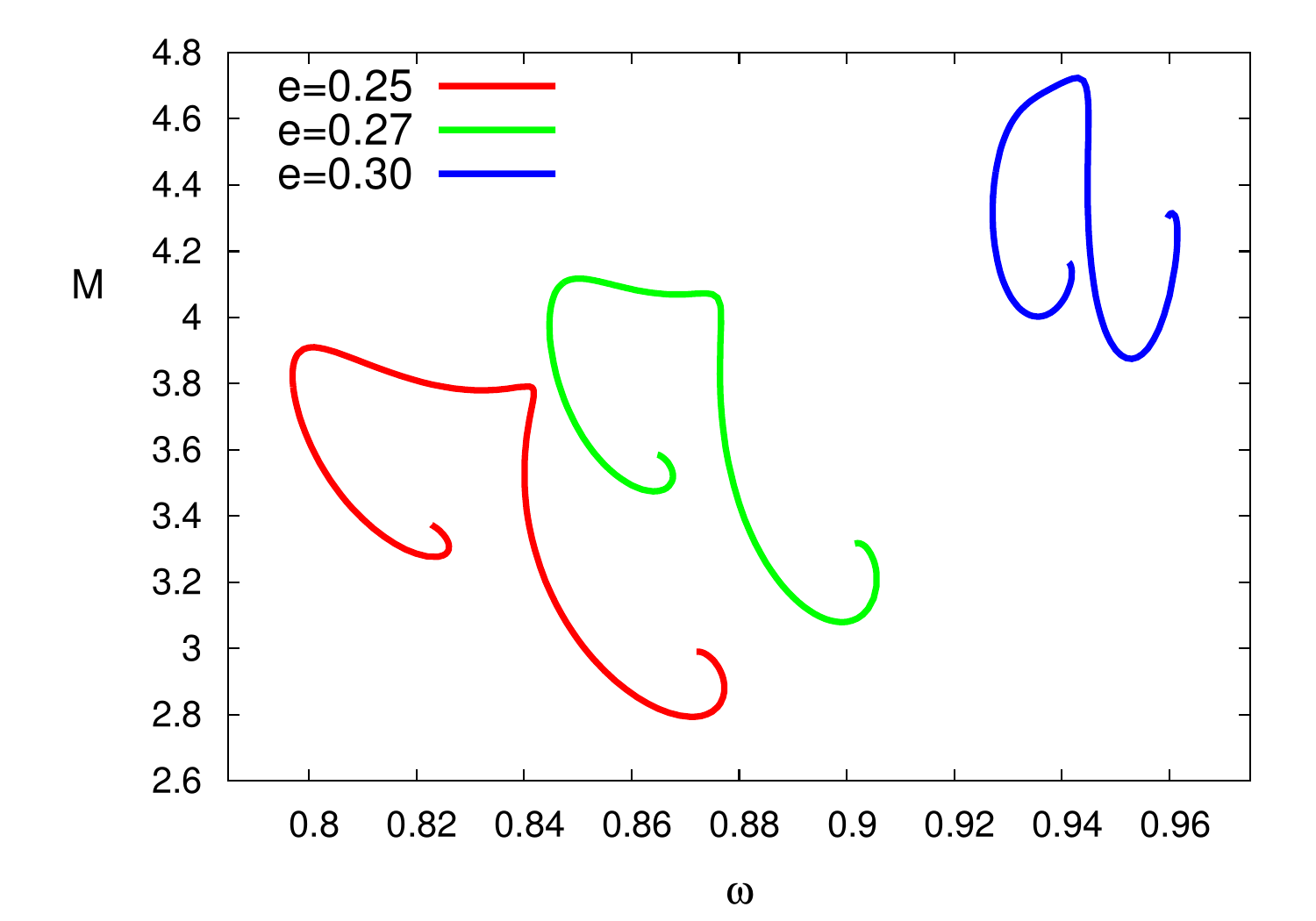}
\includegraphics[height=.32\textheight,  angle =-90]{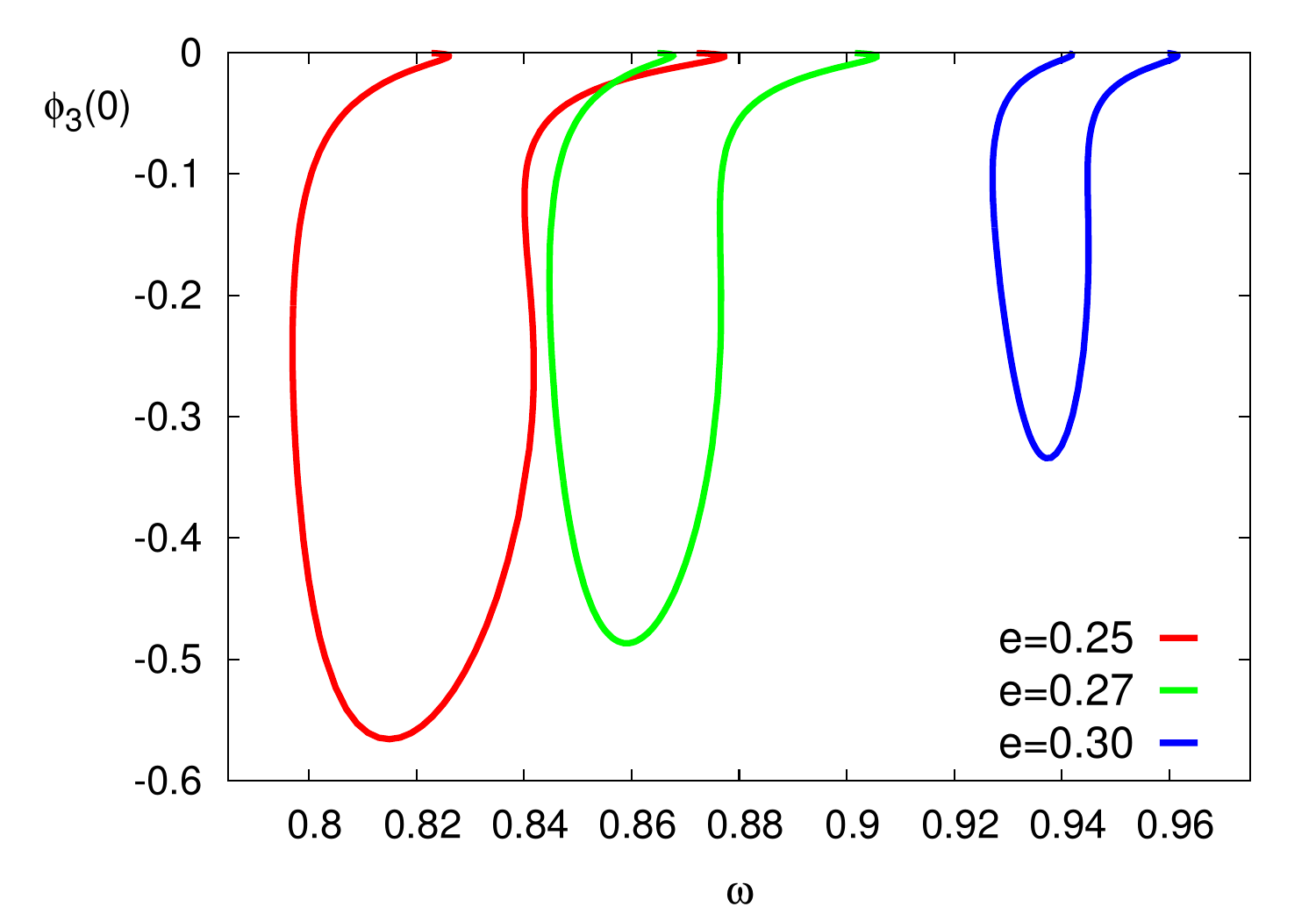}
\includegraphics[height=.32\textheight,  angle =-90]{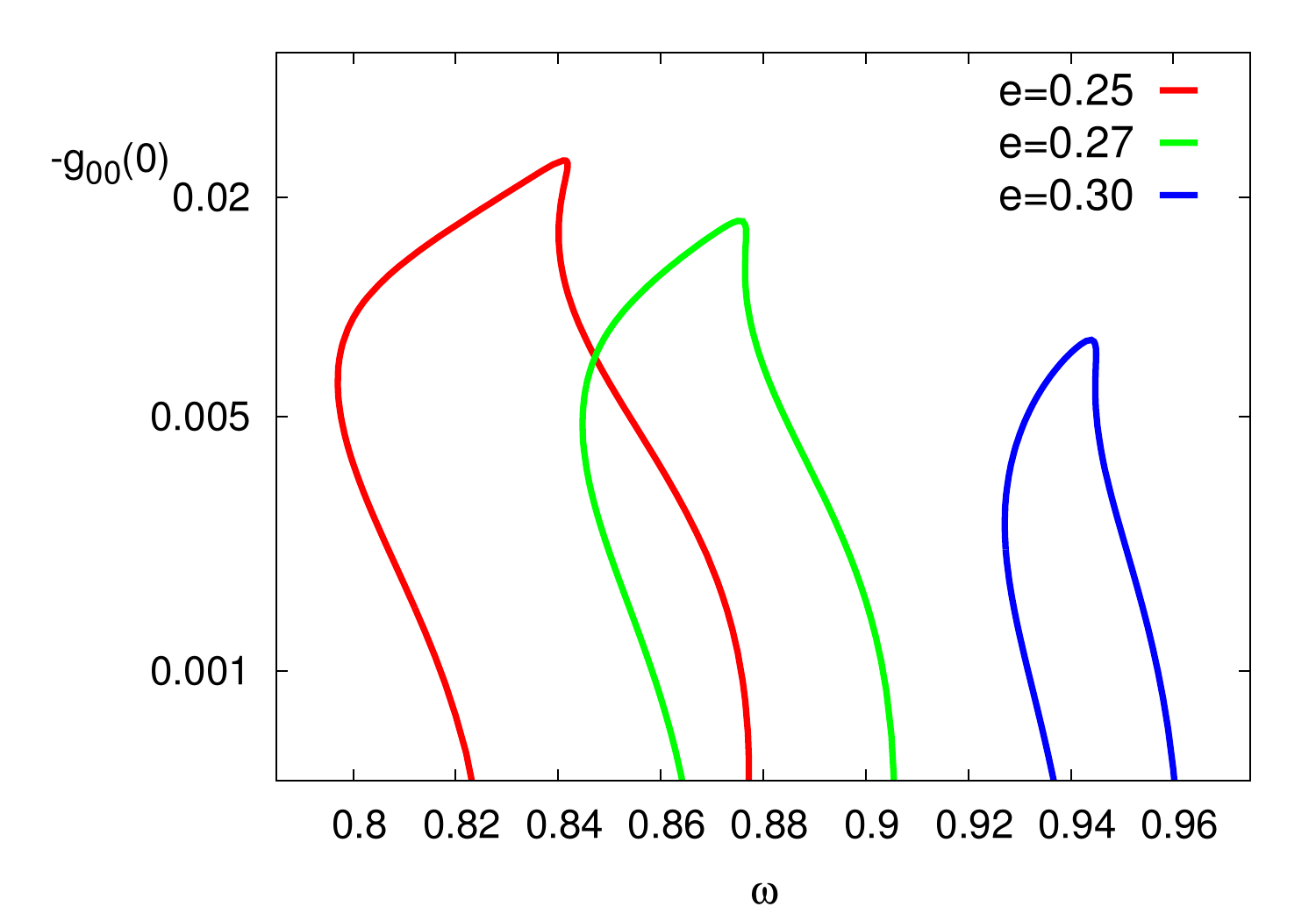}
\includegraphics[height=.32\textheight,  angle =-90]{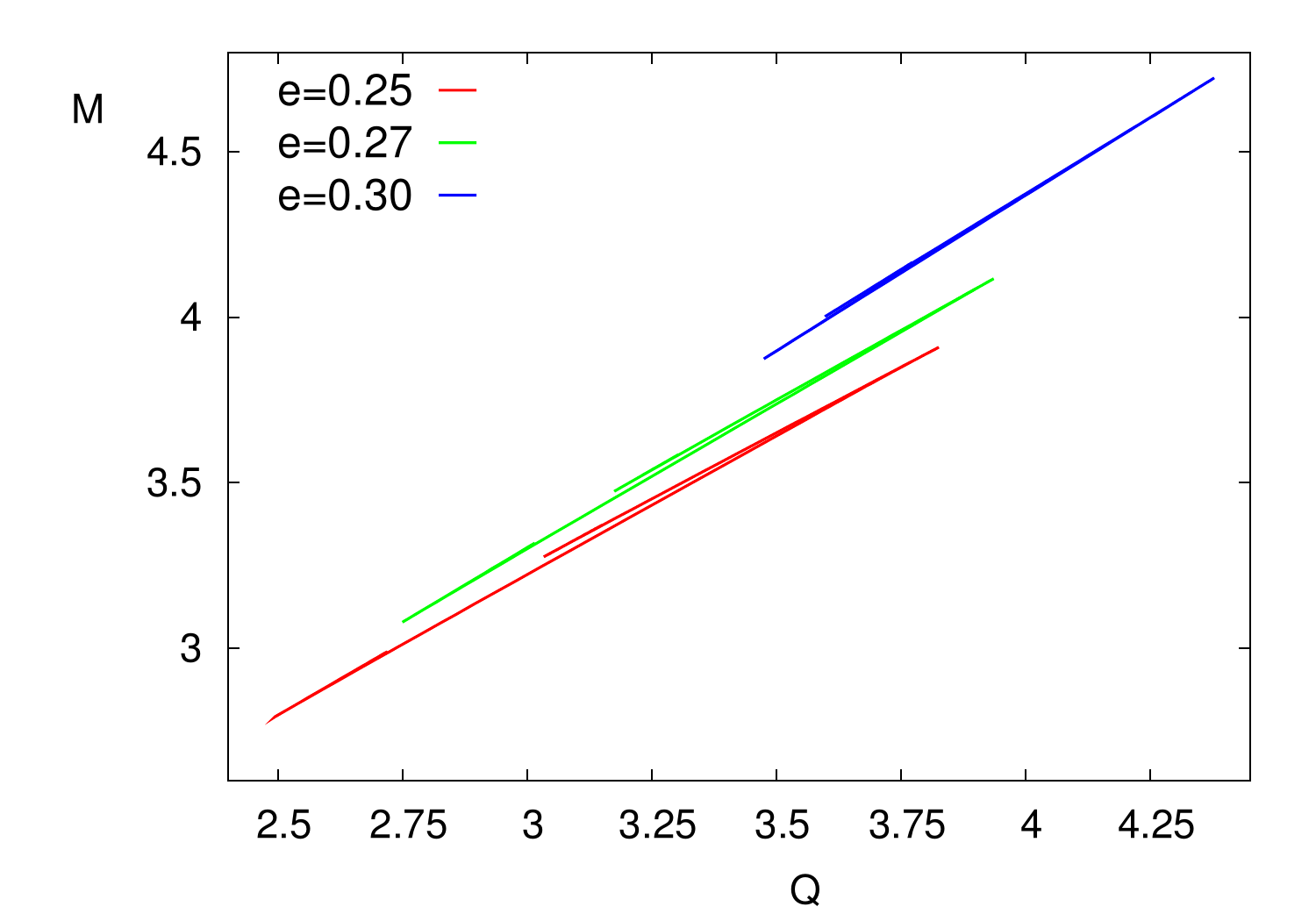}
\end{center}
\caption{\small Evolution of the detached double spirals of gauged $O(3) $ boson stars in an isolated domain: 
The ADM mass $M$ in units of $8\pi$ (upper left), the central values of the scalar component $\phi_3(0)$,(upper right), the metric component $-g_{00}(0)$ (bottom left) vs the frequency $\omega$, and the ADM mass $M$  vs the Noether charge $Q$, both in units of $8\pi$ (bottom right), are plotted for some set of values of the gauge coupling $e$ for $\alpha=0.6$ and $\beta=0.5$.}    
\lbfig{fig10}
\end{figure}

The evolution toward the bifurcation follows basically the same scenario as described above.
Figure \ref{fig9} clearly illustrates that the bifurcation is happening between $e=0.20$ and $e=0.25$ for $\alpha=0.6$.
As in the ungauged case, the spirals of both types of solutions here disconnect from their branches at the bifurcation and reconnect with each other.
The evolution of the detached double spirals is illustrated further in Fig.~\ref{fig10}.
As before, the remaining branches reconnect and form a connected set of boson stars covering the full $\omega$ domain.

Second, given a value of $\alpha$, there may be a certain value of the gauge coupling $e$, for which the competing forces of the attractive gravitational potential and the long-range electrostatic interaction are balanced at the mass threshold $\omega=1$. 
This then allows for a new pattern to occur, similar to the one seen for the usual Q-balls in Minkowski spacetime \cite{Rosen:1968mfz,Friedberg:1976me,Coleman:1985ki}, where both the energy and the charge of the solutions diverge as $\omega \to 1$. 
In such a case, the limit $\omega \to 1$ is no longer Newtonian. 
This pattern is indicated by the $e=0.3$ curves in Fig.~\ref{fig9}.
In contrast to the global charges, the central values of the electrostatic potential $A_0(0)$ and the metric function $g_{00}$ remain finite in this limit. 

The two bottom plots in Fig.~\ref{fig9} exhibit the $M(Q)$ curves of the solutions.
Here we also indicate (with black lines) the mass of $Q$ free scalar quanta $M = \mu Q$.
Configurations of type I solutions are expected to be stable below the line, and to become quantum mechanically metastable when crossing the line while retaining classical stability up to the cusp, assuming the usual stability considerations apply \cite{Friedberg:1986tq}.

A final twist is encountered as the electric coupling $e$ is increased above another critical value.
Here the type I $O(3)$ boson stars cease to exist.
The evolution toward this critical value of $e$ is also seen in the mass-frequency plot of Fig.~\ref{fig9}.
As seen in the figure, this will happen slightly above $e=0.30$.
The physical mechanism behind this pattern is likely analogous to the one seen for the usual $U(1)$ gauged boson stars \cite{Jetzer:1989us,Jetzer:1992tog,Pugliese:2013gsa,Kleihaus:2009kr,Kumar:2014kna,Kunz:2023qfg}: 
the electrostatic repulsion becomes stronger than the gravitational attraction.\par

\begin{figure}[t!]
\begin{center}
\mbox{ \hspace*{-1.0cm}
\includegraphics[height=.32\textheight,  angle =-90]{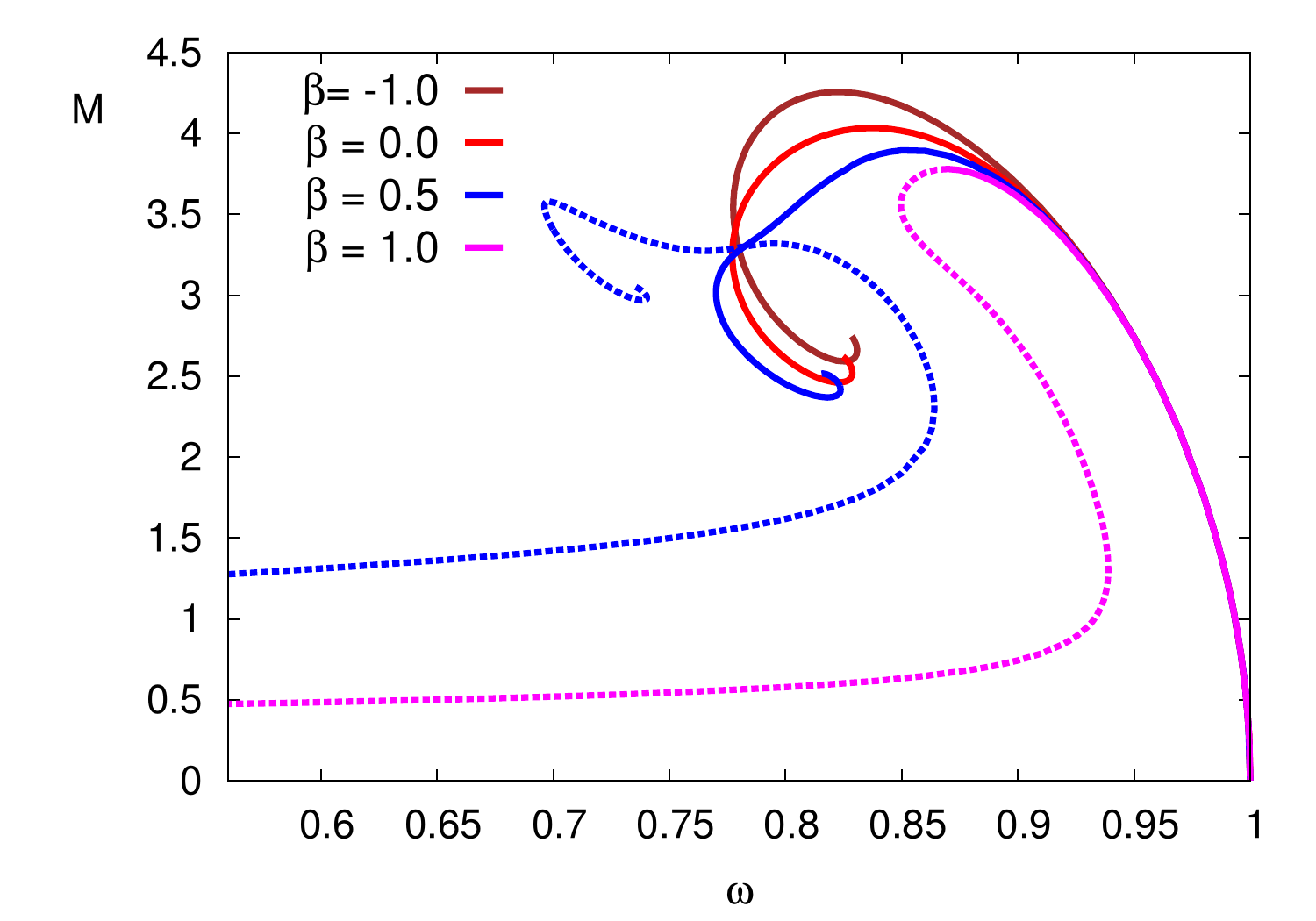}
\includegraphics[height=.32\textheight,  angle =-90]{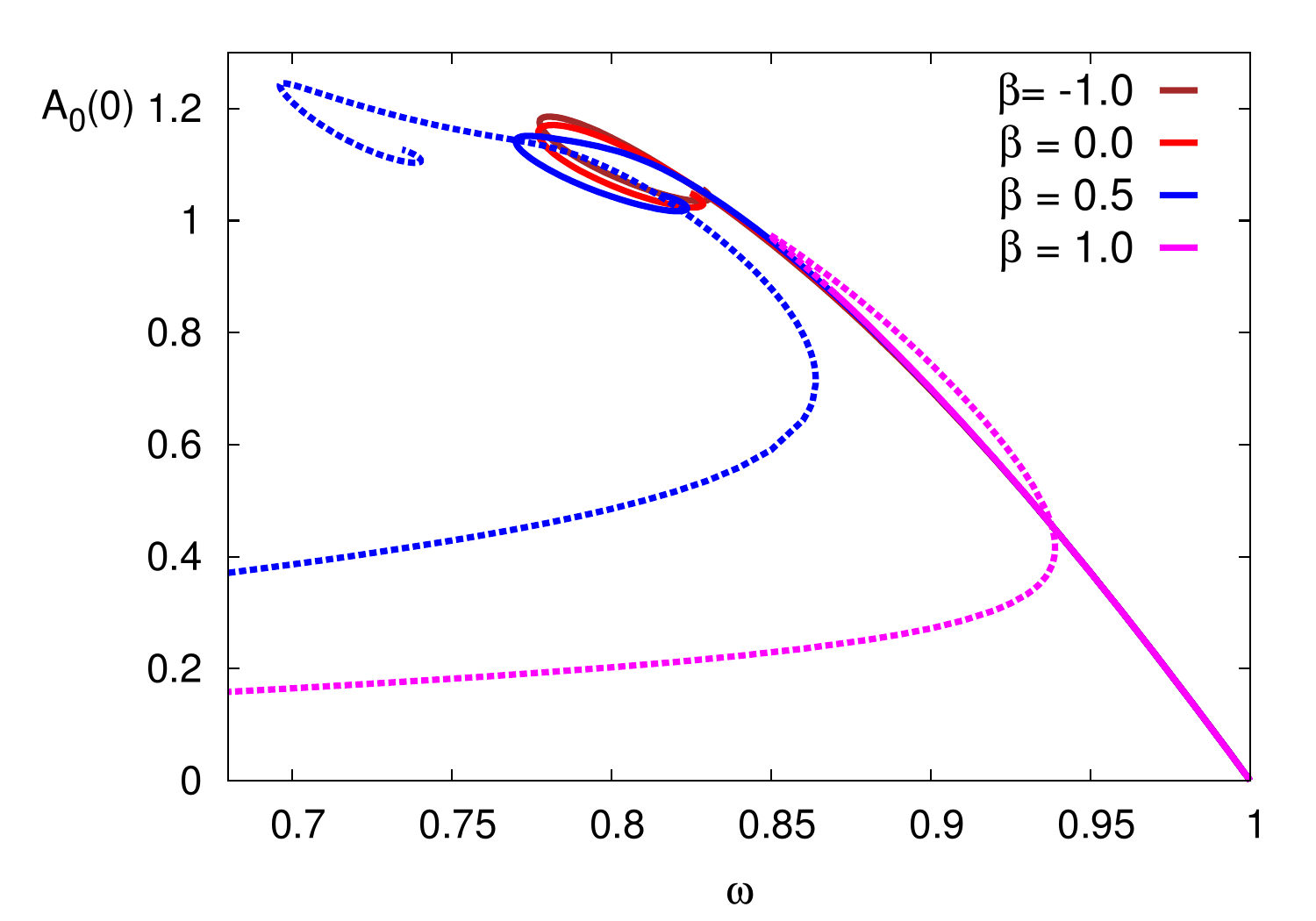}
}
\end{center}
\caption{\small Spherically symmetric $U(1)$ gauged  $O(3)$ boson stars: 
The ADM mass $M$ in units of $8\pi$ (left plot) and 
the central values of the electrostatic potential $A_0$ (right plot) 
vs the frequency $\omega$ are plotted for $\alpha=0.6$  and $e=0.2$  for some set of values of the rescaled parameter $\beta$.
The solid and dashed lines correspond to solutions of types I and II, respectively.}
\lbfig{fig11}
\end{figure}

\begin{figure}[t!]
\begin{center}
\includegraphics[height=.285\textheight,  angle =0]{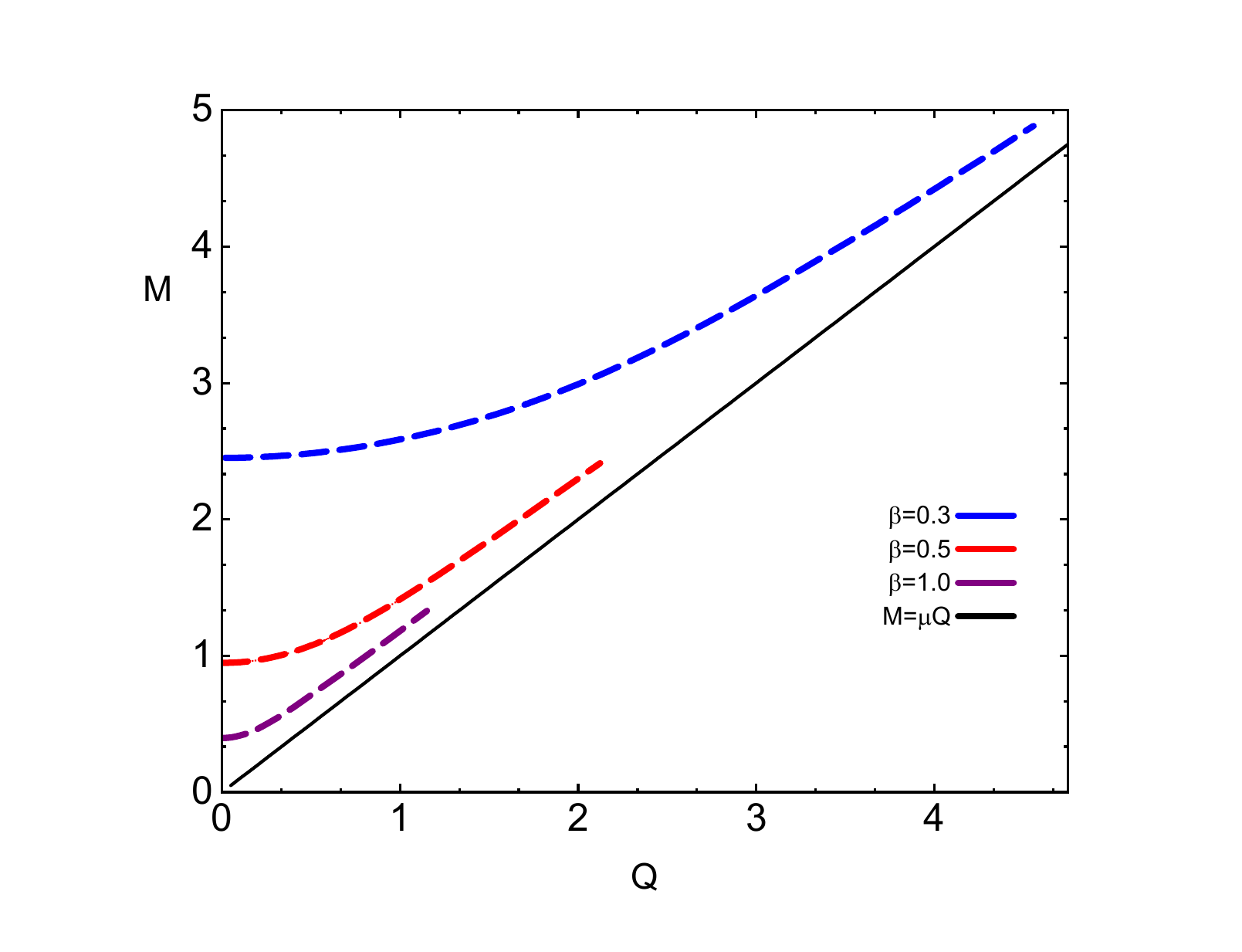}
\includegraphics[height=.278\textheight,  angle =0]{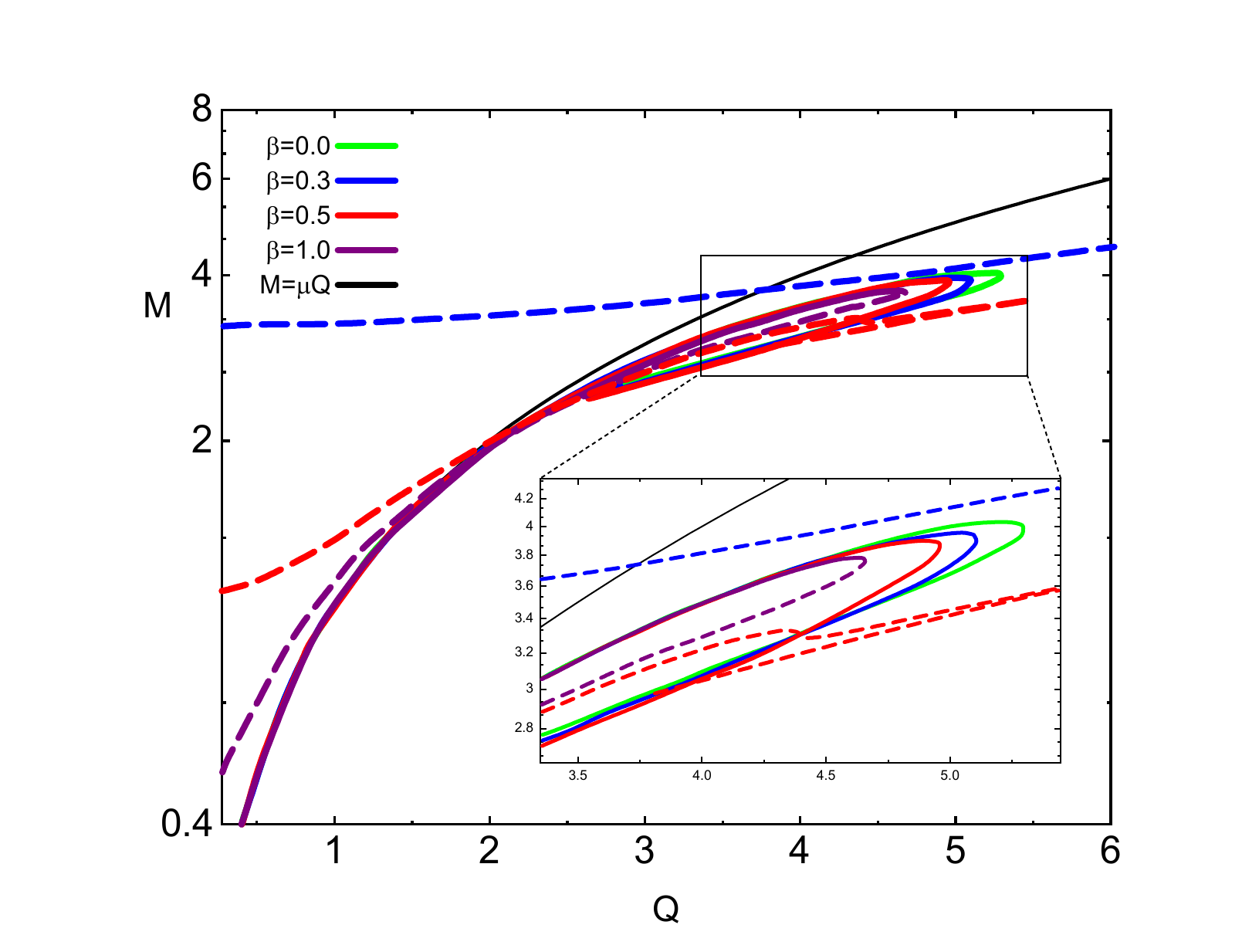}
\end{center}
\caption{\small Spherically symmetric $U(1)$ gauged $O(3)$ boson stars: 
The ADM mass $M$ of type I and type II boson stars vs the Noether charge $Q$, both in units of $8\pi$, for $\alpha=0.3$ (left plot) and for $\alpha=0.6$ (right plot), $e=0.2$, and some set of values of the $\beta$. The solid and dashed lines correspond to solutions of types I and II, respectively. The line $M=\mu Q$ (black) indicates the mass of Q free scalar quanta.}    
\lbfig{fig12}
\end{figure}

Also for the $U(1)$-gauged solutions, it is interesting to consider the evolution with $\beta$ as shown in Figs.~\ref{fig11} and \ref{fig12} for $e = 0.2$. 
Now the electrostatic repulsion supplements the scalar interaction in balancing the gravitational attraction, leading to higher masses of the charged solutions as seen by comparing Fig.~\ref{fig11} (left plot) with Fig.~\ref{fig8} (right plot).   
While Fig.~\ref{fig12} (right plot) illustrates the dependence of the mass on the charge for non-negative values of $\beta$, Fig.~\ref{fig11} (left plot) also includes  a negative value of $\beta$.
As discussed above, for $\beta=0$ and negative values of $\beta$ type II solutions do not form since they need a negative potential.

\begin{figure}[t!]
\begin{center}
\includegraphics[height=.285\textheight,  angle =0]{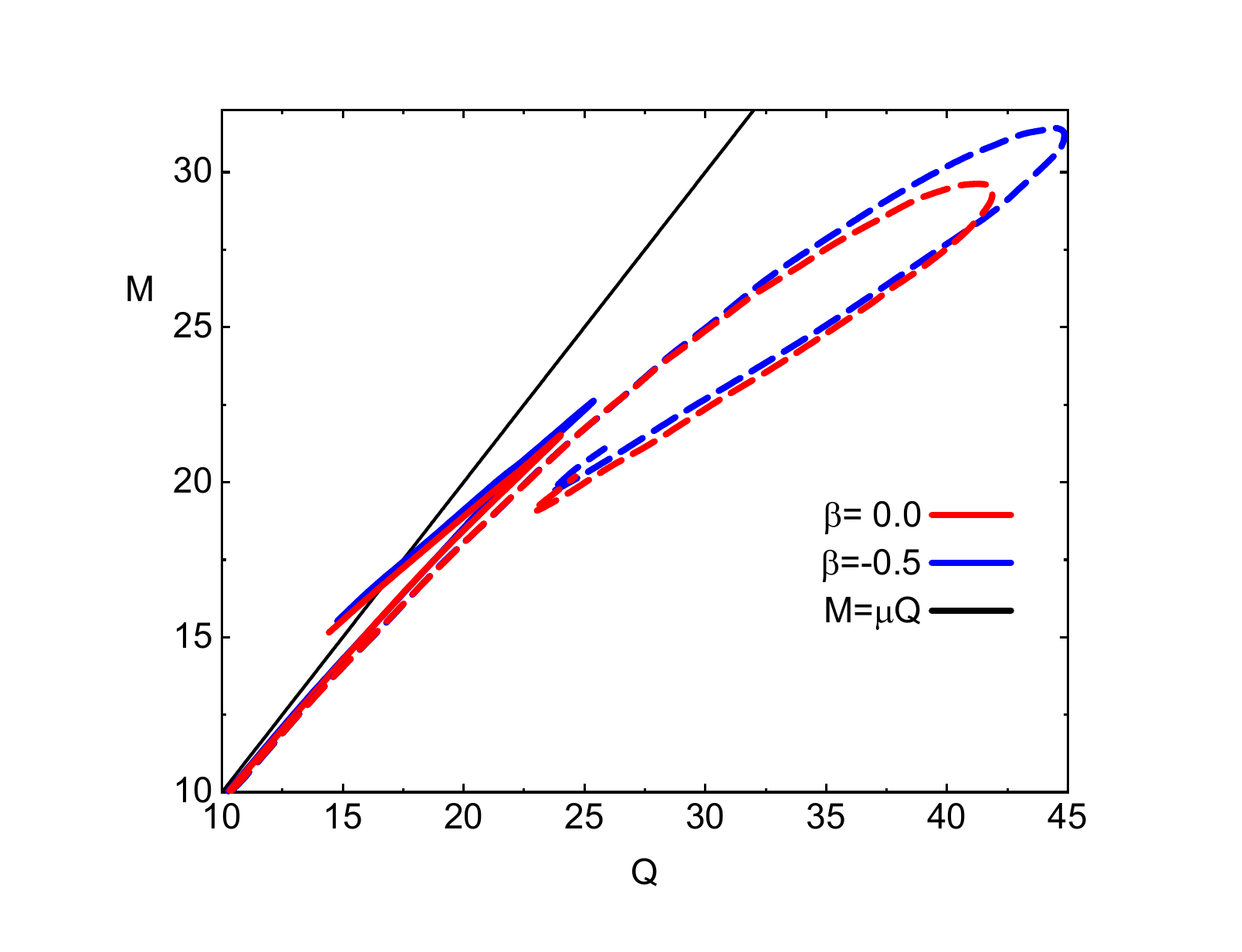}
\includegraphics[height=.285\textheight,  angle =0]{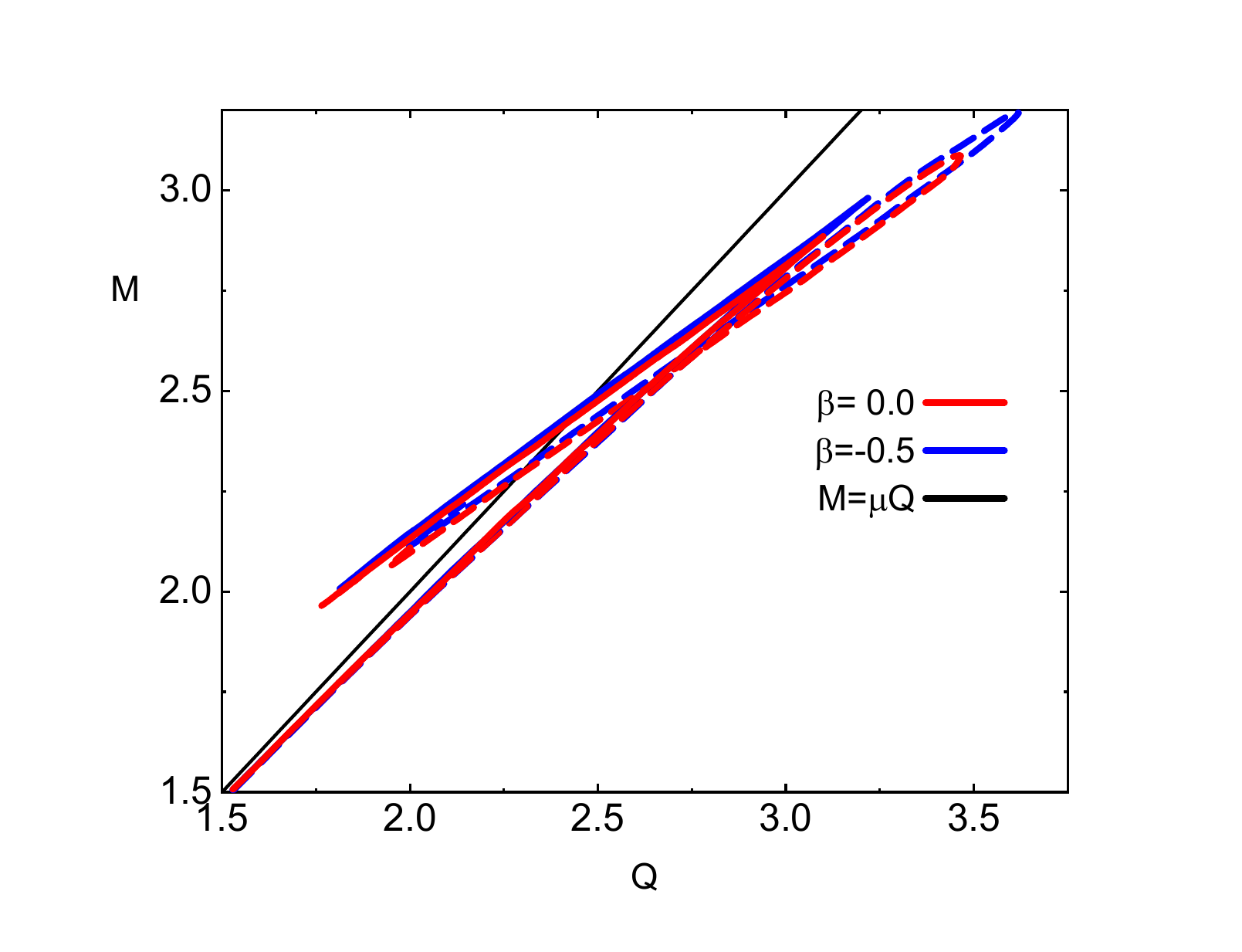}
\includegraphics[height=.285\textheight,  angle =0]{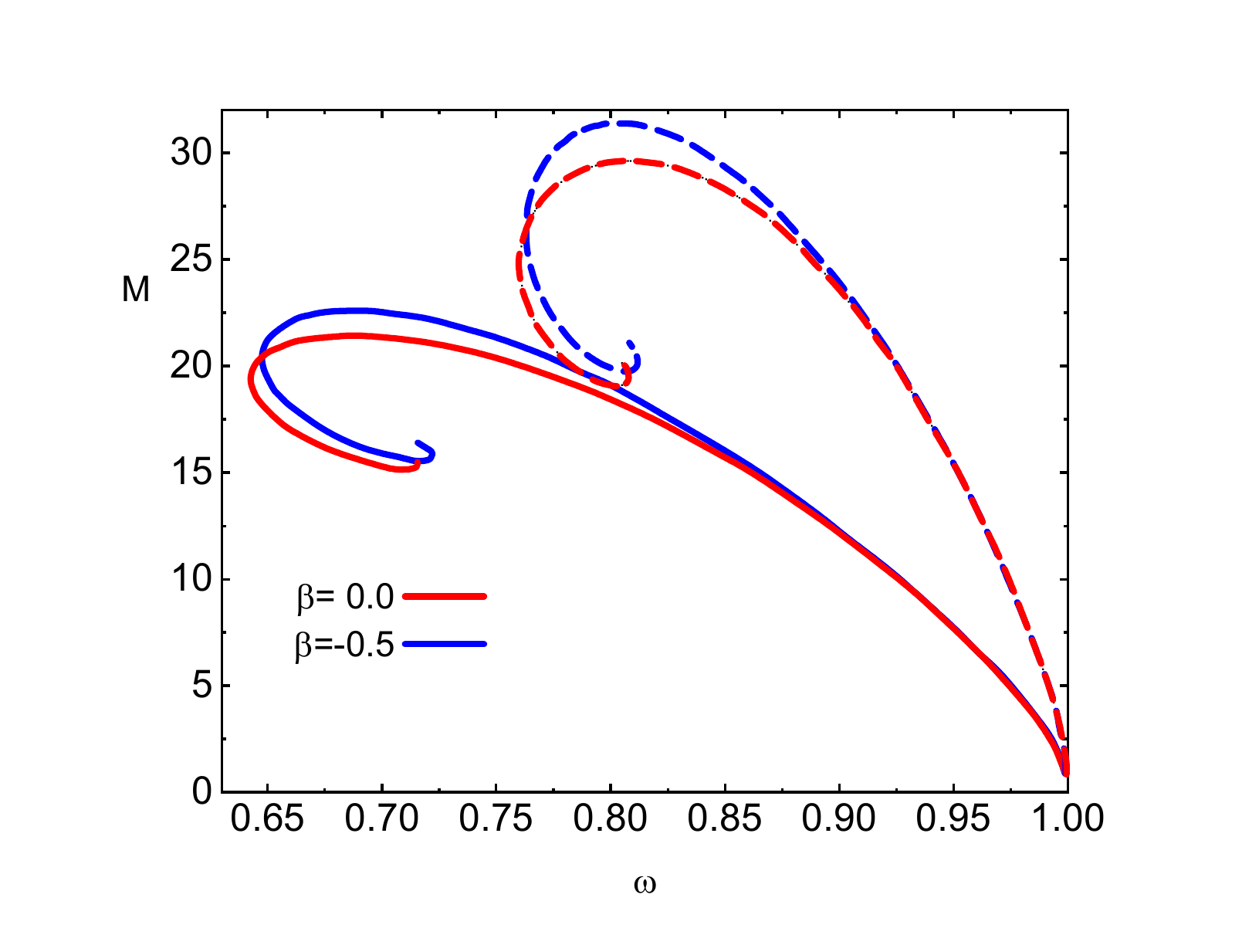}
\includegraphics[height=.285\textheight,  angle =0]{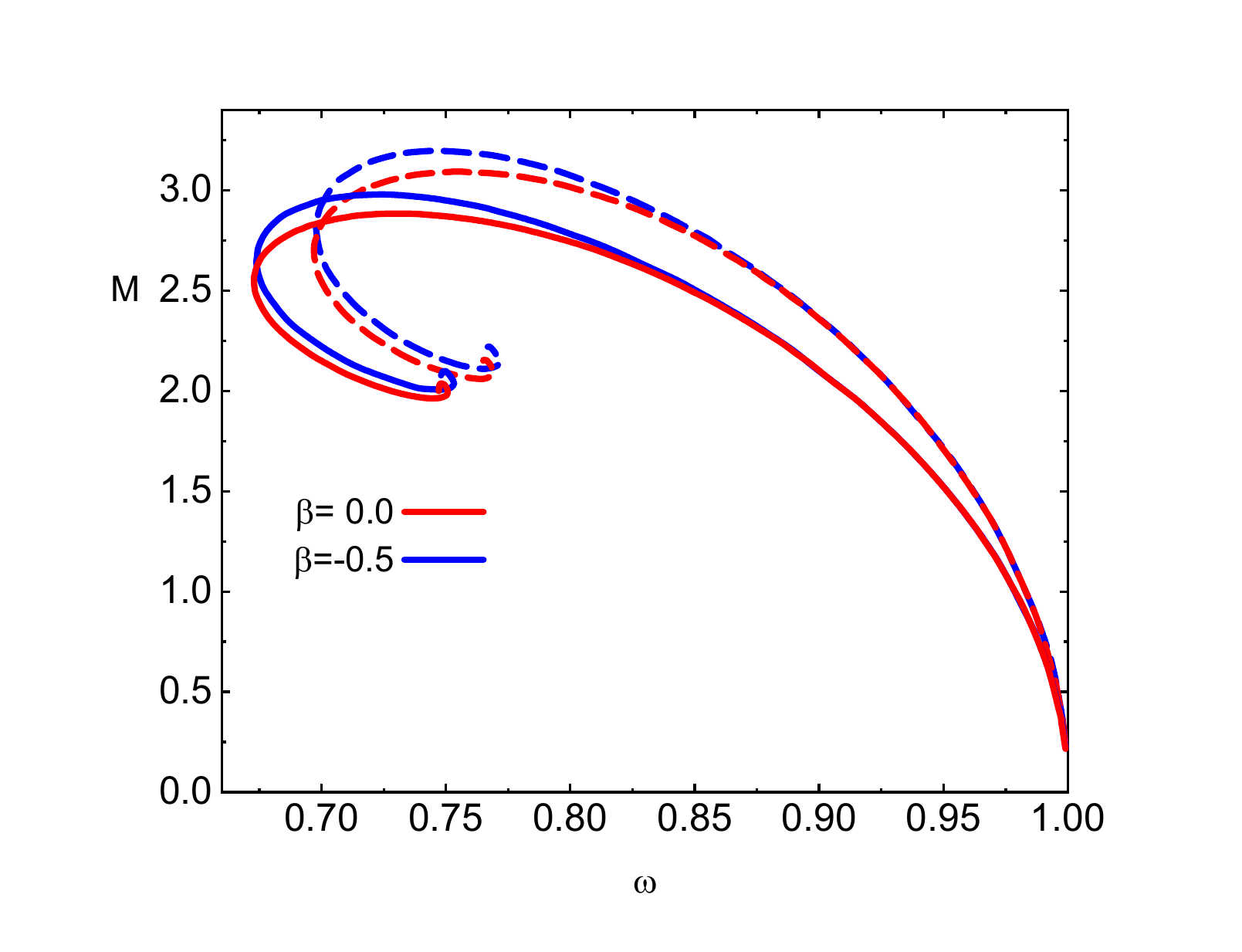}
\end{center}
\caption{\small Spherically symmetric gauged and ungauged $O(3)$ boson stars: 
The ADM mass $M$ of type I  boson stars  vs the Noether charge $Q$, both  in units of $8\pi$, for $\beta=0$ and $\beta=-0.5$ and for $\alpha=0.3$ (upper left), for $\alpha=0.6$ (upper right) and the ADM mass $M$ of type I  boson stars in units of $8\pi$ vs the frequency $\omega$ for $\alpha=0.3$ (bottom left), for $\alpha=0.6$ (bottom right).
The solid and dashed lines correspond to ungauged and gauged $(e=0.1)$ solutions, respectively. The line $M=\mu Q$ (black) indicates the mass of Q free scalar quanta.}    
\lbfig{fig13}
\end{figure}

Figure \ref{fig12} (left plot) shows the mass versus the charge for the smaller value of the gravitational coupling $\alpha = 0.3$.
In this case there is not enough attraction compared to the electrostatic repulsion for type I solutions to exist.
The type I solutions then trivialize for any value of the $\beta$.
In contrast, the type II solutions exist within the full interval $0 \le \omega \le 1$ for $\beta>1/8$,
while their mass and charge decrease with increasing $\beta$.

Finally, we compare in Fig.~\ref{fig13} the charged ($e=0.1$) and uncharged ($e=0$) type I solutions with ($\beta=-0.5$) and without ($\beta=0$) a scalar self-interaction for two values of the gravitational coupling.
Here the solid lines indicate the ungauged boson stars, whereas the dashed lines show the gauged boson stars.
The figure shows the shift of $\omega_{\text{min}}$ of type I solutions to larger values with decreasing $\beta$, resulting in a faster trivialization of the solutions as the gauge coupling $e$ is increased.

\section{Conclusions and outlook}

In this work, we have investigated spherically symmetric, stationary boson stars in an $O(3)$ non-linear sigma model with a particular symmetry-breaking self-interaction potential \cite{Ferreira:2025xey}, extending the landscape of compact objects supported by self-gravitating scalar fields. 
Unlike the commonly studied boson star models, this framework allows for boson stars of two types.
Type I boson stars emerge as usual from the vacuum as the boson frequency $\omega$ is decreased below its upper limit, the boson mass $\mu$.
In contrast, type II boson stars emerge from a static scalaron solution, where the boson frequency vanishes, $\omega=0$ \cite{Ferreira:2025xey,Nucamendi:1995ex,Kleihaus:2013tba,Chew:2024bec}. 
At a critical value of the gravitational coupling $\alpha$ the different types of solutions bifurcate.
The spirals disconnect from their branches and reconnect with each other, while the remaining branches also reconnect and form a connected set of boson stars that extends over the full range of frequencies, $0 \le \omega < \mu$.
Thus, the model yields a rich spectrum of field configurations.

We have numerically constructed these spherically symmetric, stationary boson star solutions in both the ungauged and the gauged model.
As expected, the charged boson stars show analogous features as the uncharged ones.
The bifurcation of both types of solutions then arises at a critical value of the gauge coupling $e_{cr}$, that depends on the gravitational coupling $\alpha$.
However, at large electric coupling, type I boson stars cease to exist, and type II boson stars extend over the full frequency range.

Looking ahead, several promising avenues for further study emerge. 
While most studies of boson stars assume spherical or axial symmetry to simplify the equations of motion, relaxing these symmetry constraints opens the door to a much broader and potentially richer class of solutions \cite{Herdeiro:2020kvf,Mikhaliuk:2025mxy}. 
Boson stars without rotational symmetry -- such as those with multipolar structures -- could exhibit novel properties. 
Exploring such configurations in the above $O(3)$ model would require fully three-dimensional numerical simulations. 
Such objects might possibly occur as transient states during scalar field collapse.

A natural next step will be to analyze rotating boson stars in the above $O(3)$ framework to see how the observed pattern of solutions generalizes to the rotating case \cite{Kleihaus:2005me,Mikhaliuk:2025mxy}.
This then immediately suggests an even more interesting extension of this work that involves the study of black hole solutions endowed with bosonic hair in the context of the above $O(3)$ model \cite{Herdeiro:2014goa,Kleihaus:2015iea,Herdeiro:2018djx}. 
Such solutions challenge the classical no-hair theorems by allowing stationary scalar field configurations to persist outside the event horizon \cite{Herdeiro:2015waa}. 
Investigating the existence, stability, and observational signatures of these hairy black holes can, in principle, yield insights for future observations of gravitational waves or observations within the electromagnetic spectrum.
Indeed, as gravitational wave and multi-messenger astronomy continue to advance, boson stars and hairy black holes might represent interesting candidates in the broader quest for exotic compact objects beyond the standard black hole paradigm.

\section*{Acknowledgment}
We would like to thank Luiz Agostinho Ferreira and Eugen Radu for enlightening discussions. 
Y.S.~gratefully acknowledges support by FAPESP, project No 2024/01704-6, and the Instituto de F\'isica de S\~ao Carlos, IFSC, for kind hospitality during the initial stage of this work.

\begin{small}

\end{small}
\end{document}